\documentclass[final,a4paper]{gGAF2e}
\usepackage[utf8x]{inputenc}
\usepackage{tikz}
\usepackage[normalem]{ulem}
\usepackage[a4paper,colorlinks=true,breaklinks=true]{hyperref}
\usepackage{color}
\usepackage{natbib}
\usepackage{amsmath,amsfonts,amssymb}
\usepackage{graphicx}
\usepackage[mathlines]{lineno}
\bibpunct[; ]{(}{)}{,}{a}{}{;}
\usepackage{doi}
\usepackage{url}

\newcommand{\Sc}{\ensuremath{\mathrm{Sc}}}
\newcommand{\Pra}{\ensuremath{\mathrm{Pr}}}
\newcommand{\Pplus}{\ensuremath{\mathrm{P_+}}}
\newcommand{\Pminus}{\ensuremath{\mathrm{P_-}}}
\newcommand{\Rt}{\ensuremath{\mathrm{R_t}}}
\newcommand{\Rc}{\ensuremath{\mathrm{R_c}}}
\newcommand{\Ra}{\ensuremath{\mathrm{Ra}}}
\newcommand{\Le}{\ensuremath{\mathrm{Le}}}
\newcommand{\rr}{\ensuremath{{\hat{\bm{r}}}}}
\newcommand{\pos}{\ensuremath{{\bm{r}}}}
\newcommand{\kk}{\ensuremath{{\hat{\bm{k}}}}}
\renewcommand{\Theta}{{\varTheta}}
\renewcommand{\Gamma}{{\varGamma}}
\renewcommand{\Xi}{{\varXi}}
\renewcommand{\Phi}{{\varPhi}}
\renewcommand{\Psi}{{\varPsi}}
\renewcommand{\Omega}{{\varOmega}}
\usepackage{cancel}
\allowdisplaybreaks
\reversemarginpar
\usepackage[percent]{overpic}
\usepackage{color}

\title{The onset of thermo-compositional convection in rotating spherical shells}

\author
{LUIS SILVA$^1$, JAMES F.~MATHER$^2$ and RADOSTIN D.~SIMITEV$^3$
\thanks{$^1$ \href{mailto:lacsilva@gmail.com}{lacsilva@gmail.com}, \url{orcid.org/0000-0003-3421-0009}}
\thanks{$^2$ \url{orcid.org/0000-0001-9518-0515}}
\thanks{$^3$
  \href{mailto:Radostin.Simitev@glasgow.ac.uk}{Radostin.Simitev@glasgow.ac.uk},  \url{orcid.org/0000-0002-2207-5789}} \\\vspace{6pt}
School of Mathematics and Statistics, University of Glasgow -- Glasgow
G12 8QQ, UK\\ 
\received{Received by GAFD on 2018-07-11; Accepted by GAFD on 2019-06-28} 
}

\begin{document}

\jvol{xx} \jnum{xx} \jyear{2018} \jmonth{}
\markboth{Silva, Mather \& Simitev}{Onset of thermo-compositional convection}


\maketitle

\begin{abstract}
Double-diffusive convection driven by both thermal and compositional
buoyancy in a rotating spherical shell can exhibit a rather large
number of behaviours often distinct from that of the single
diffusive system. In order to understand how the differences in
thermal and compositional molecular diffusivities determine the
dynamics of thermo-compositional convection we investigate
numerically the linear onset of convective instability in a
double-diffusive setup. 
We construct an alternative equivalent formulation of the
non-dimensional equations where the linearised double-diffusive
problem is described by an effective Rayleigh number, $\Ra$, measuring
the amplitude of the combined buoyancy driving, and a second
parameter, $\alpha$, measuring the mixing of the thermal and
compositional contributions. This formulation is useful in that it
allows for the analysis of several limiting cases and reveals 
dynamical similarities in the parameters space which are not obvious
otherwise.
We analyse the structure of the critical curves in this $\Ra-\alpha$
space, explaining asymptotic behaviours in $\alpha$, transitions
between inertial and diffusive regimes, and transitions between large
scale (fast drift) and small scale (slow drift) convection. We perform
this analysis for a variety of diffusivities, rotation rates and shell
aspect ratios showing where and when new modes of convection take place.
\end{abstract}


\section{Introduction}

Convection in a rotating spherical fluid shell provides one of the
fundamental models for understanding the large-scale
motions and the magnetic fields observed in many geophysical, planetary and
astrophysical systems, see \citep{Jones2011,glatzmaier2013,Busse2015}.
{Thermal and compositional convection occur, for instance, in
Earth's outer core which} is composed mostly of iron and nickel,
alloyed to lighter elements, supposedly silicon, sulphur and oxygen 
\citep{JEANLOZ1990}. Heat is continually lost to outer space
{establishing a secular-cooling} thermal gradient that can drive
thermal convection in the outer core. In addition, secular cooling
leads to freezing of iron onto the inner core, a process in which both
latent heat and light material are released and give rise 
to additional thermal and chemical buoyancy \citep{JACOBS1953,BRAGINSKY1963}. 
Both buoyancy components are important when modelling the geodynamo.
Indeed, recent estimates of thermal conductivity for iron at core
conditions \citep{PozzoEtAl2012,DaviesEtAl2015} confirm that outer
core convection cannot be driven by thermal buoyancy alone. It is
estimated that the compositional contribution to the buoyancy flux in
the Earth's core is around 80\% \citep{LISTER199517}. 
To model thermo-compositional convection \citet{BraginskyRoberts95}
suggested that temperature and
concentration can be combined into a single ``co-density''
field.  The co-density formulation has since been widely used in
numerical simulations of the geodynamo and planetary dynamos, see
reviews \citep{Jones2011,Christensen2015}. However, the co-density
formulation requires that a single set of effective boundary 
conditions, a single effective distribution of sources and a single
effective value of the {co-density} diffusivity must be
used. {While the co-density formalism has been adapted to 
mimic thermo-compositional boundary conditions in the Core \citep{Hori2012},
stratification \citep{Olson2017,Takehiro2018}, the Ganymede snow zone 
\citep{Christensen2015a} and other situations, it is nor clear how well
the it captures the dynamics of the distinct heat and composition fields.}
Only few {modelling studies of core dynamics} have employed a
genuine 'double-diffusive' formulation where both temperature and
composition are included as separate fields.  \citet{BreuerEtAl2010}
found that the convective flow patterns differ significantly depending
on the dominant driving component with abrupt changes in differential
rotation, mean kinetic 
energy and transport efficiencies indicating a regime change when the
relative contribution of compositional driving exceeds
20\%. \citet{TrumperEtAl2012} extended the latter work and concluded
that the main reason for this flow behaviour is the pronounced
difference in diffusivities as measured by the Prandtl numbers of the
thermal component and the compositional component. \citet{TrumperEtAl2012}
also considered the effect of distinct boundary conditions for
temperature and concentration and obtained some preliminary results on
the onset of convection using an initial value code. Some comparison
with \citep{TrumperEtAl2012} is made in our article. In a
double-diffusive model of Mercury's dynamo \citet{ManglikEtAl2010}
observed that when thermal and compositional buoyancy are of equal
intensity, finger convection penetrates the upper layer enhancing the
poloidal magnetic field, a significant difference compared to
co-density cases. Exploring a geodynamo model \citet{Takahashi2014}
confirmed that the morphology of the poloidal field is determined by
the balance of thermal and compositional driving with a dipolar
magnetic field maintained when thermal buoyancy comprises less than
60\% of the total driving and non-dipolar fields otherwise prevailing
due to  helicity reduction and concluded that the fraction of power
injection by thermal convection in the present geodynamo is below this
threshold. 

A systematic study of the linear onset of double-diffusive
convection {in rotating spherical shells is required} for
understanding the behaviour of turbulent dynamos driven by thermal and
compositional  buoyancy. Indeed, it is often found that the properties
of convection 
at onset provide much insight to finite-amplitude convection
\citep{Simitev2003,Simitev2006b}. 
{Much is known about the onset of purely thermal convection --
we refer to the recent monograph of \citet{Zhang2017} which provides
an extensive list of references.} 
{Interest in double-diffusive convection has been motivated
mostly by oceanic applications. The most fascinating feature of
double-diffusive convection is that instability can occur in a fluid
where the vertical density gradient is statically stable. Surprisingly, 
this instability is due to diffusion, a typically stabilising effect.
When the instability is driven by the small-diffusivity component it is
called fingering; when the instability driven by the large-diffusivity
component it is known as diffusive convection; see \citep{Radko2013}.
The linear onset in a two-component fluid layer
constrained between two horizontal boundaries was studied by
\citep{Stern1960,Nield1967,Veronis1968,Baines1969} who naturally
focussed on the fingering and diffusive regimes. The effect of 
rotation with respect to a vertical axis was considered by
\citet{Pearlstein1981} who found that a non-rotating layer can be
destabilized by rotation, and that a rotating layer can be
destabilized by the addition of a bottom-heavy solute gradient. These
results were summarized in the context of core convection by
\citet{Fearn1988}.  
The effect of rotation perpendicular to gravity and the effect of
inclined top and bottom boundaries is crucial in our case. 
These effects were considered by \citet{Busse2002} and by \citet{Simitev2011}
in the setting of a rapidly rotating cylindrical annulus with conical
caps. \citet{Busse2002} established that when rotation and gravity are
mutually perpendicular the second buoyancy component can balance
the Coriolis force and so significantly facilitate convection.} 
\citet{Simitev2011} found that due to additional
‘‘double-diffusive’’ eigenmodes, the neutral curves for the onset of
instability in a rotating annulus are typically multi-valued and form
regions of instability in the parameter space that may be entirely
disconnected from each other. It was also observed that while known
asymptotic expressions for the critical Rayleigh number and frequency
derived by \citet{Busse2002} describe the onset of convection over an
extended range of non-asymptotic parameter values but do not capture
the full complexity of the critical curves.  
\cite{Net2012} studied numerically the influence of an externally
enforced compositional gradient on the onset of convection of a
mixture of two components in a rotating fluid spherical shell. Both
positive and negative compositional gradients were considered in the
latter study and it was found that the influence of the mixture is
significant in both cases, in particular the critical values of the
thermal Rayleigh number, of the frequency and of the wave number
depends strongly on the direction of the compositional gradient.
The aim of the present article is to extend and complement the latter
studies and to contribute to a detailed linear analysis of
double-diffusive convection in rotating spherical shells
{including fingering, diffusive as well as direct top-heavy
regimes}. Indeed, there is much to learn since both  
the parameter space and the space of valid modelling assumptions that
can be made is very large. For instance, our model is mathematically
similar to those 
of \cite{Busse2002} and \cite{Simitev2011} but is set in a spherical
geometry. In contrast to \cite{Net2012}, we consider the stress-free
case for the velocity boundary conditions and internal rather than
differential heating. Similarly to \cite{Net2012} we 
allow for both a stabilizing compositional gradient which may occur for
instance, in lower main-sequence stars with heated helium-rich core
surrounded by lighter hydrogen layers \citep{Kippenhahn2012}, as well as
destabilising compositional gradients relevant, for instance, in the
in the case of the Earth's core where solidification with the release of
light components takes place. In extension to \cite{Net2012} we
consider negative values of the thermal expansion coefficient a situation relevant in  some cases
\citep{SquyresEtAl83,RottgerEtAl1994}. In comparison to the linear analysis of
\citep{TrumperEtAl2012} we use an eigenvalue solver that offers many
additional details on the onset of convective instability.

The article is organised as follows. We start by reviewing the mathematical
set-up used to estimate the parameters of the system at onset in
section~\ref{s:mathematicalFormulation}. In section \ref{s:EffRaFormalism} we
introduce a formalism that allows us to measure the combined effect of
the thermal and  compositional buoyancy and to better understand the
convective processes at play. Section \ref{s:phenomenology} proceeds to describe
the competition of eigenmodes that leads to the formation of the
global critical curves for onset. 
The remaining sections~\ref{s:PrDependence}, \ref{s:tauDependence}
and \ref{s:etaDependence} are devoted to describing the onset of
convection depending on the Prandtl and Coriolis numbers and aspect
ratio, respectively. A summary of the results and conclusions is
presented in section~\ref{s:conclusions}.

\section{Mathematical formulation and numerical solution}
\label{s:mathematicalFormulation}
\label{s:problemSetup}
We investigate the onset of thermo-compositional convection in a
rotating spherical shell. The shell has a thickness $d=r_o-r_i$, where 
$r_o$ and $r_i$ are the inner and the outer radii,
respectively, and rotates about an axis aligned with the $z$-direction
at a constant rate $\Omega$. The unit vectors pointing in the
$z$-direction and in the radial direction  are denoted by {$\kk$ and
$\rr$, respectively, and $\pos$ is the position vector with
  magnitude $r$}. 
{We assume that the spherical shell is full of} incompressible
fluid  solution with constant kinematic viscosity $\nu$, thermal
diffusivity $\kappa$, and chemical diffusivity $D$. 
{We take the} density $\rho$ of the fluid to depend linearly on 
changes in composition and temperature with first order expansion
coefficients $\alpha_C$ and $\alpha_T$, respectively. We
employ the Boussinesq approximation {in that} variations in density
are assumed important only when {affecting} the gravitational force
$-\rho\gamma \pos$, with $\gamma$ a constant. 
In order to {isolate the effects induced by differences
in thermal and chemical diffusivities} we follow
\citep{Busse2002,Net2012} and disregard any differences in source-sink
distribution and in boundary conditions for the temperature $T$ and
the concentration $C$. Static profiles $T(r)$ and $C(r)$ with radial
gradients $\upartial_r T =-\beta_T r$ and $\upartial_r C=-\beta_C r$
then exist {assuming the} temperature and concentration are fixed at the
boundaries and have  uniformly distributed sources with constant
densities $\beta_T$ and $\beta_C$, respectively.

Using  $d^2/\nu$ as the unit of time, $d$ as the unit of length,
$T^*=\beta_T d^2\nu/\kappa$ as the unit of temperature and $C^*=\beta_C
d^2\nu/D$ as the {unit} of concentration {we} arrive at the
following linearised equations in adimensional units, 
\begin{subequations}
\label{eq:convection}
\begin{gather}
\label{eq:NavierStokes}
\upartial_t \bm{u} =
- \tau \kk \times \bm{u}
- {\bm{\nabla}} \pi
+ (\Rt \Theta +\Rc \chi )\pos
+ \nabla^2 \bm{u}, \\
{{\bm{\nabla}} {\bm{\cdot}} \bm{u} = 0,}
\\
\label{eq:temperature}
   \upartial_t \Theta = \Pra^{-1}\nabla^2 \Theta - \bm{u} {\bm{\cdot}} {\bm{\nabla}} T , \\
\label{eq:composition}
   \upartial_t \chi =   \Sc^{-1} \nabla^2 \chi - \bm{u} {\bm{\cdot}} {\bm{\nabla}} C,
\end{gather}
\end{subequations}
where $\bm{u}$ is the flow velocity, $\pi$ is an {effective pressure
  including all terms that can be written in gradient form}, {and
$\Theta$ and $\chi$ are the temperature and the compositional
anomalies} from the static reference states $T$ and $C$, respectively. The non-dimensional
parameters are defined in Table~\ref{t:modelParameters}.  
Note that here the {compositional} scale, $C^*$, is inversely
proportional to $D$ in contrast with the work of \citet{Simitev2011}
who scales $C^*$ with $\kappa$. As a consequence,  
the compositional Rayleigh numbers {reported bellow} must be
{divided by} the \emph{Lewis} number, $\Le=\kappa/D$, when compared
to the latter work. 
\begin{table}
\begin{center}
\begin{tabular}{l c}
\hline\\[-2mm]
\textbf{Parameter} & \textbf{Definition}\\[2mm]
\hline \\[-3mm]
Coriolis Number        & $\tau = {2}\Omega d^2/\nu$ \\[1mm]
Thermal Rayleigh Number       & $\Rt = \alpha_T \gamma d^4 T^*/\nu^2$
\\[1mm]
Compositional Rayleigh Number & $\Rc = \alpha_C \gamma d^4 C^*/\nu^2$
\\[1mm]
Thermal Prandtl Number        & $\Pra = \nu/\kappa$  \\[1mm]
Compositional Prandtl Number (Schmidt Number) & $\Sc = \nu/D$  \\[1mm]
Radius Ratio                  & $\eta=r_i/r_o$ \\[1mm]
\hline
\end{tabular}\\[1mm]
\caption{Non-dimensional model parameters. {In the text we
    occasionally refer to both the Prandtl and the Schmidt numbers
    as ``Prandtl numbers'' for brevity.} 
}  \label{t:modelParameters}
\end{center}
\end{table}
Except when otherwise mentioned, the equations are solved for a
spherical shell with an  inner to outer radius ratio of $\eta=0.35$. 

{Exploiting the solenoidality of the velocity field we use}
the poloidal-toroidal decomposition
\begin{align}
\bm{u}={\bm u}_P+{\bm u}_T={\bm{\nabla}}\times{\bm{\nabla}}\times {\cal S}(\pos,t){\bm r}+{\bm{\nabla}}\times {\cal T}(\pos,t){\bm r},
\end{align}
{and represent $\bm{u}$ in terms of a poloidal and a toroidal scalar
functions ${\cal S}(\pos,t)$ and ${\cal T}(\pos,t)$, respectively.}
{All unknown scalar quantities, ${\cal X}(\pos,t)\equiv[{\cal S},
  {\cal T}, \Theta,\chi]^\top(\pos, t)$ are then} assumed to obey a linear Fourier mode ansatz in time 
\begin{equation*}
  {\cal X}(\pos,t)=    \tilde {\cal X}(\pos)\; \exp({\mathrm{i}}t(\omega-{\mathrm{i}}\Gamma)),
\end{equation*}
where $\omega$ is the frequency of oscillation (or drift rate), $\Gamma$ is the
growth rate, {and $^\top$ denotes transpose}. When  $\Gamma$ is negative
perturbations decay and the system is considered stable, otherwise
instability occurs. 
Operating on equation~\eqref{eq:NavierStokes} by 
${\pos}{\bm{\cdot}}{\bm{\nabla}}\times$ and by ${\pos}{\bm{\cdot}}{\bm{\nabla}}\times{\bm{\nabla}}
\times$ 
four scalar equations are obtained
\begin{subequations}
\begin{gather}
\label{eq:Poloidal}
\hspace*{-25em}({\mathrm{i}}\omega +\Gamma) (\nabla^2-2r\upartial_r)(\nabla^2_{H} \tilde {\cal S}) 
\\
\hspace{4em}
=- \tau \pos {\bm{\cdot}} {\bm{\nabla}\times} {\bm{\nabla}\times}(\kk \times \tilde{\bm{u}})
+ \nabla^2_{H}(\Rt \tilde\Theta + \Rc \tilde\chi ) + \pos {\bm{\cdot}} {\bm{\nabla}\times} ({\bm{\nabla}\times}(\nabla^2 \tilde{\bm{u}})), \nonumber\\
\label{eq:Toroidal}
({\mathrm{i}}\omega +\Gamma) (\nabla^2_{H} \tilde {\cal T}) =
- \tau \pos {\bm{\cdot}} {\bm{\nabla}\times} ( \kk \times \tilde{\bm{u}})
+ \pos {\bm{\cdot}} {\bm{\nabla}\times} (\nabla^2 \tilde{\bm{u}}), \\
\label{eq:temp}
({\mathrm{i}}\omega +\Gamma) \tilde \Theta =
  \Pra^{-1}\nabla^2 \tilde\Theta
- \tilde{\bm{u}}{\bm{\cdot}}{\bm{\nabla}} T , \\
\label{eq:comp}
({\mathrm{i}}\omega +\Gamma) \tilde \chi =
  \Sc^{-1}\nabla^2 \tilde\chi
- \tilde{\bm{u}}{\bm{\cdot}} {\bm{\nabla}} C ,
\end{gather}
\label{eq:equations}
where the horizontal Laplacian 
\begin{equation*}
 \nabla^2_{H} f = \nabla^2 f - \frac{1}{r^2}\frac{\upartial\ }{\upartial r}\left(r^2 \frac{\upartial f}{\upartial r} \right),
\end{equation*}
is introduced and some terms with $\tilde{\bm{u}}$ are {left
unexpanded for brevity}. 
{A the spherical boundaries temperature and composition anomalies
  are assumed to vanish and stress-fee velocity boundary conditions are imposed}
{\begin{gather}
\tilde{\cal S} = \upartial_{rr} \tilde{\cal S} = \upartial_{r}(\tilde{\cal T}/r)
  =\tilde\Theta=\tilde\chi  = 0 \quad \text{at} \quad  r = r_i, r_o.
\end{gather}}
\end{subequations}

{In order} to find the locus of marginal convective stability where
$\Gamma=0$ the  scalar quantities $\tilde {\cal X}(\pos)$ are further expanded in
terms of spherical harmonics for the angular part. Due to the
linearity of the equations and of the orthogonality properties of
spherical harmonics, individual azimuthal wave numbers $m$ decouple
and can be investigated one by one. To complete the spatial
discretisation of the problem in the radial direction we follow
\citep{ZhangBusse87,ArdesEtAl97} and expand $\tilde {\cal X}(\pos)$ in
trigonometric functions obeying the boundary conditions. After
computing the appropriate Galerkin projection integrals numerically
equations \eqref{eq:equations} take the matrix form 
\begin{equation}
({\mathrm{i}}\omega +\Gamma)
 \left[ A \right]_{n,l} 
 \begin{bmatrix}
    \tilde s_{n,l,m}\\
    \tilde t_{n,l,m}\\
    \tilde \theta_{n,l,m}\\
    \tilde \chi_{n,l,m}
\end{bmatrix}
  =
 \left[ B \right]_{n,l}
        \begin{bmatrix}
        \tilde s_{n,l,m}\\
        \tilde t_{n,l,m}\\
        \tilde \theta_{n,l,m}\\
        \tilde \chi_{n,l,m}
        \end{bmatrix},
\label{eq:algebraic}
\end{equation}
for fixed $m$, with summation implied over the degree of the
associated Legendre polynomials $l$ and the index of the radial functions $n$.
Matrices $A$ and $B$ are of the form
\begin{equation}
\left[ A \right]_{n,l} =
\begin{bmatrix}
\square     & \varnothing & \varnothing & \varnothing \\
\varnothing & \square     & \varnothing & \varnothing \\
\varnothing & \varnothing & \square     & \varnothing \\
\varnothing & \varnothing & \varnothing & \square \\
\end{bmatrix}
, \quad
\left[ B \right]_{n,l} =
\begin{bmatrix}
\square     & \square     & \square     & \square     \\
\square     & \square     & \varnothing & \varnothing \\
\square     & \varnothing & \square     & \varnothing \\
\square     & \varnothing & \varnothing & \square     \\
\end{bmatrix},
\end{equation}
where squares represent non-null blocks. A triangular truncation of
the sums is chosen such that there are the same number of radial functions as
associated Legendre polynomials \citep{ZhangBusse87,ArdesEtAl97},
\begin{equation}
2n+l-m+2 \leq 3 +2N,
\end{equation}
with $N$ an integer bigger than 2. $N$ represents the required resolution for
the calculation and has been set to values higher than 10 and as close
as feasibly possible to $m$. Equation~\eqref{eq:algebraic} is then
solved for the complex eigenvalues $({\mathrm{i}}\omega +\Gamma)$ using standard 
numerical eigenvalue methods implemented in the LAPACK
library\footnote{The LAPACK Library, Linear Algebra PACKage
 (LAPACK) \url{http://www.netlib.org/lapack}.}. 

Once the eigenvalue problem \eqref{eq:algebraic} is solved for a set of
fixed parameter values non-trivial numerical extremization and
continuation problems must be tackled in order to follow the marginal
stability curves in the parameter space. The numerical code for the
solution of the problem can be obtained via \citep{silva2018a}.

\section{An effective Rayleigh-number formalism}
\label{s:EffRaFormalism}

\begin{figure}
{\footnotesize (a) \hspace{0.45\textwidth} (b)}\\
\includegraphics[height=5.5cm,width=\columnwidth]{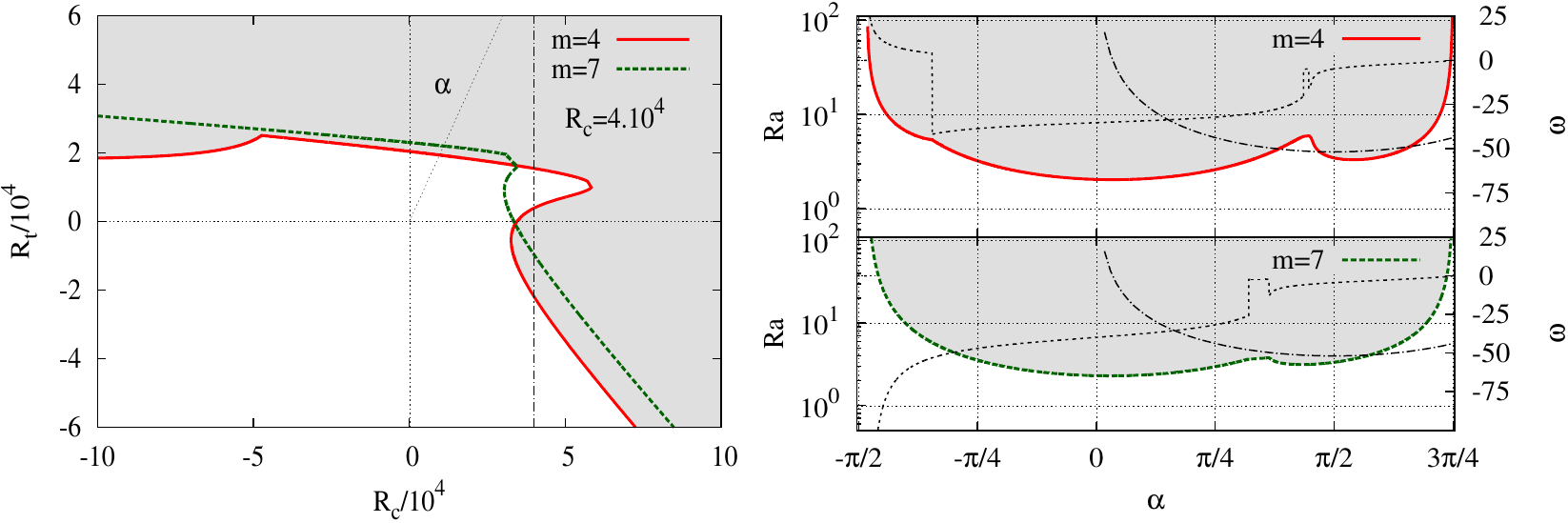}
	\caption{(a) {Composite critical curves and drift rates
          for {selected} wave numbers in the $\Rc-\Rt$ plane (a)
          and in the $\alpha-\Ra$ plane (b). Parameter values are
          $m=4$ (red solid curve) and $m=7$ (green dashed curve) at
          $\tau=1.2\times10^3$, $\Pra=1$, $\Sc=10$ and $\eta=0.35$. Drift rates
          $\omega$ are plotted in panels (b) using thin black dashed
          lines and are measured on the right $y$-axes. In all panels
          the dashed-dotted curve has equation $\Rc=4\times10^4$ and
          is used to denote a section along which plots for figure~\ref{f:eigenModes}
          are made.} The shaded areas denote the 
                regions where convective instability occurs. (Color online)}
	\label{f:IntroducingRcRtplane}
\end{figure}

Multivalued critical curves for the onset of thermo-compositional
convection in related problems were found in
\citep{Simitev2011,Net2012} {and occur} in the present setting
{as well} as illustrated in figure~\ref{f:IntroducingRcRtplane}. 
{It is difficult to find the extrema of multivalued curves and to
  this end we introduce below a new adimensional parameter
$\Ra$ in terms of which the critical curves become single valued.
This new parameter} can be used to  measure criticality with respect to
the onset of convection similarly to the purely thermal case.
{An added advantage is that} the associated transformation of the
{linearised} governing equations {makes it easy to interpret}
two important limiting cases as discussed within this section. 
{To introduce $\Ra$ we note that} the momentum equation
\eqref{eq:NavierStokes} can be transformed by representing the
buoyancy force in the form  
\begin{equation}
\left(\Rt\ \Theta + \Rc\ \chi\right)\pos =
\Ra\, \big(\cos(\alpha) \Theta + \sin(\alpha) \chi\big)\pos,
\end{equation}
{with $\alpha$ varying between $-\pi/2$ and $3\pi/4$} and $\Ra$
being positive. {We refer to $\Ra$ as \emph{the effective Rayleigh
number}. It is} related to the thermal and the compositional Rayleigh numbers by 
\begin{equation}
\Ra = \sqrt{\Rt^2 + \Rc^2}.
\end{equation}
We refer to $\alpha$ {as} the \emph{Rayleigh mixing angle} since it
corresponds to an angle in the \Rt-\Rc~ plane. It is {also defined
in terms of} the thermal and the compositional Rayleigh numbers,
$$
\alpha=\text{atan2}{(\Rc ,\Rt)},
$$
where the function $\text{atan2}(y,x)$ is {the four-quadrant
inverse tangent} defined for $x\in\mathbb{R}$, $y\in \mathbb{R}$ as the principal
argument $\text{Arg}(z)$ of the complex number $z=x+iy$, a notation
used in many programming languages.  
The case of purely thermal convection corresponds to  $\alpha=0$, the
case of purely compositional convection is obtained at $\alpha=\pi/2$
and the typical co-density approach is corresponds to $\alpha=\pi/4$ with $\Pra=\Sc$.

We also remark that our parametrisation has several
advantages over previous parametrisations such as the once proposed by
\cite{BreuerEtAl2010} and \citet{TrumperEtAl2012}.
These authors considered a total  Rayleigh number constructed
{simply} as the sum of the thermal and compositional  Rayleigh
numbers. This is a very good approximation for the cases when the Prandtl and the
Schmidt numbers are equal $\Pra = \Sc$ but fails when they are even
only marginally different. 
{In addition to converting critical curves to single valued functions,
and providing a single positive measure of criticality, our
reparametrisation, maps transitions between convective regimes and
large Rayleigh number regions to asymptotes in the $\alpha$.}
In more practical terms, plotting on logarithmic scales can be
employed due to the introduction of the strictly positive effective
Rayleigh number. 

{The Rayleigh angle $\alpha$ introduced above appears in some
respects similar to the Turner angle $\text{Tu}$ defined by 
\citet{Ruddick1983} but the two are not directly related as they arise
in different models of double-diffusive convection. 
$\text{Tu}$ is used in the oceanographic literature to  determine the
local stability of an inviscid saline water column
\citep{McDougall1988} and arises as a parameter in the so called
``unbounded gradient layer'' model where density increases downwards
and instabilities are due to local gradients of temperature and
concentration \citep{Radko2013}. In this case double-diffusive  
convection is characterised by an intrinsic length scale which makes it
possible to non-dimensionalise this model so 
that $\text{Tu}$ (or equivalently the so called stability density
ratio $R_\rho$) is the only parameter needed to control instability
and identify ``salt-fingering'' and ``diffusive'' regimes. In
contrast, we are motivated by planetary core applications where
``bottom-light'' convection  with concurrently destabilizing
temperature and composition is arguably more important. The natural
length scale of vigorous bottom-light convection is not a function
only of the ratio of local gradients of temperature and concentration
but is also determined by other factors including geometry and
driving. To account for these we employ a ``vertically-bounded layer''
model of the type first used by \citet{Veronis1968}. Here buoyancy
forces are driven by the global variation or temperature and
concentration across the volume of the 
spherical shell and two Rayleigh numbers are now needed to describe
the convective instability. The effective Rayleigh number $\Ra$ and the
Rayleigh angle $\alpha$ are a convenient reparametrisation of those as
discussed above.  
}

{To take advantage of our reparametrisation,} we now define a new
set of dynamical variables $\Psi$ and $\Psi'$ and static reference
profiles $\Xi$ and $\Xi'$ using the transformations 
\begin{subequations}
	\label{eq:newRaFormalismVars}
	\begin{gather}
	\Psi  = \Theta \cos\alpha + \chi \sin\alpha, \qquad
	\Psi' = \Theta \cos\alpha - \chi \sin\alpha, \\
	\Xi  = T \cos\alpha + C \sin\alpha, \qquad
	\Xi' = T \cos\alpha - C \sin\alpha,
	\end{gather}
\end{subequations}
with the inverses being
	\begin{gather}
	\label{eq:oldFields}
	\Theta \cos\alpha = \frac{\Psi + \Psi'}{2}, \qquad
	\chi \sin\alpha = \frac{\Psi - \Psi'}{2}.
	\end{gather}
By substituting these expressions into equations~(\ref{eq:NavierStokes}--\ref{eq:composition}) and 
adding and subtracting equations~\eqref{eq:temperature} and \eqref{eq:composition}, 
we arrive at the transformed problem
\begin{subequations}
	\label{eq:newRaFormalism}
	\begin{gather}
	\label{eq:newRaFormalismNS}
	\upartial_t \bm{u} = - \tau \kk \times \bm{u} - {\bm{\nabla}} \pi + \Ra \Psi \pos +
	\nabla^2 \bm{u}, \\
	\label{eq:newRaFormalismRho}
	\upartial_t \Psi  = \Pplus^{-1} \nabla^2 \Psi  + \Pminus^{-1} \nabla^2 \Psi' - 
	\bm{u}
	{\bm{\cdot}} {\bm{\nabla}} \Xi , \\
	\label{eq:newRaFormalismRhoP}
	\upartial_t \Psi' = \Pplus^{-1} \nabla^2 \Psi' + \Pminus^{-1} \nabla^2 \Psi  - 
	\bm{u}
	{\bm{\cdot}} {\bm{\nabla}} \Xi',
	\end{gather}
\end{subequations}
with effective Prandtl numbers defined as 
\begin{gather}
\Pplus^{-1} = \frac{\Pra^{-1} + \Sc^{-1}}{2}, \qquad
\Pminus^{-1} = \frac{\Pra^{-1} - \Sc^{-1}}{2}.
\label{eq:NewPrandtls}
\end{gather}
Note that, while the number $\Pplus$ is strictly positive, the
number $\Pminus$ can be negative thus providing means of
concentrating rather than diffusing the fields $\Psi$ or $\Psi'$. This
mechanism can drive convection at lower than expected Rayleigh numbers
and at very large scales as we shall see.

Formulation \eqref{eq:newRaFormalism} allows us to consider two
important limiting cases. Firstly, we note that the transformed momentum
equation~\eqref{eq:newRaFormalismNS} only depends on the field $\Psi$. A
consequence of this decoupling is that, in the case of $\Pminus^{-1}$ 
approaching 0, the system will behave as if driven only by one buoyancy
generating field. This situation arises when $\Pra$ is close 
or equal to $\Sc$ and thus, we obtain the {usual} co-density approximation.
We consider this case in more depth in sections~\ref{s:samePrandtl} and
\ref{s:smallDepartures}.

A second limiting case occurs when the Prandtl numbers are
significantly different in magnitude. Then the new parameters can be approximated as 
	\begin{gather}
	\label{eq:PrWhenPt.ne.Pc}
	\Pplus^{-1} \approx     \frac{1}{2} \max(\Pra^{-1},
        \Sc^{-1}),\qquad 
	\Pminus^{-1} \approx \pm \frac{1}{2} \max(\Pra^{-1}, \Sc^{-1}),
	\end{gather}
and equations~(\ref{eq:newRaFormalism}b,c) can be rewritten approximately as
\begin{subequations}
	\label{eq:rhoWhenPt.nePctot}
	\begin{gather}
	\label{eq:rhoWhenPt.ne.Pc}
	\upartial_t \Psi \approx \frac{\max(\Pra^{-1}, \Sc^{-1})}{2} \nabla^2 (\Psi  \pm \Psi') - \bm{u} 
	{\bm{\cdot}}
	{\bm{\nabla}} \Xi , \\
	\label{eq:rhopWhenPt.ne.Pc}
	\upartial_t \Psi' \approx \frac{\max(\Pra^{-1}, \Sc^{-1})}{2} \nabla^2 (\Psi' \pm \Psi ) - \bm{u} {\bm{\cdot}}
	{\bm{\nabla}} \Xi',
	\end{gather}
\end{subequations}
with the symbol $\pm$ taking the positive sign when $\Pra^{-1} \gg
\Sc^{-1}$, and the negative sign otherwise.
With all other parameters fixed, the changes to the critical value of $\Ra$ can
be due only to different behaviours of the system with respect to $\alpha$.
We consider this case in more detail in section~\ref{s:largePrDifferences}
where we will also explore the effects of $\Pra$ on the curves $\Ra_c(\alpha)$.

\section{Construction of global critical curves}
\label{s:phenomenology}
\begin{figure}
{\footnotesize (a) \hspace{0.48\textwidth} (b)} \\
\includegraphics[height=5.3cm,width=\columnwidth]{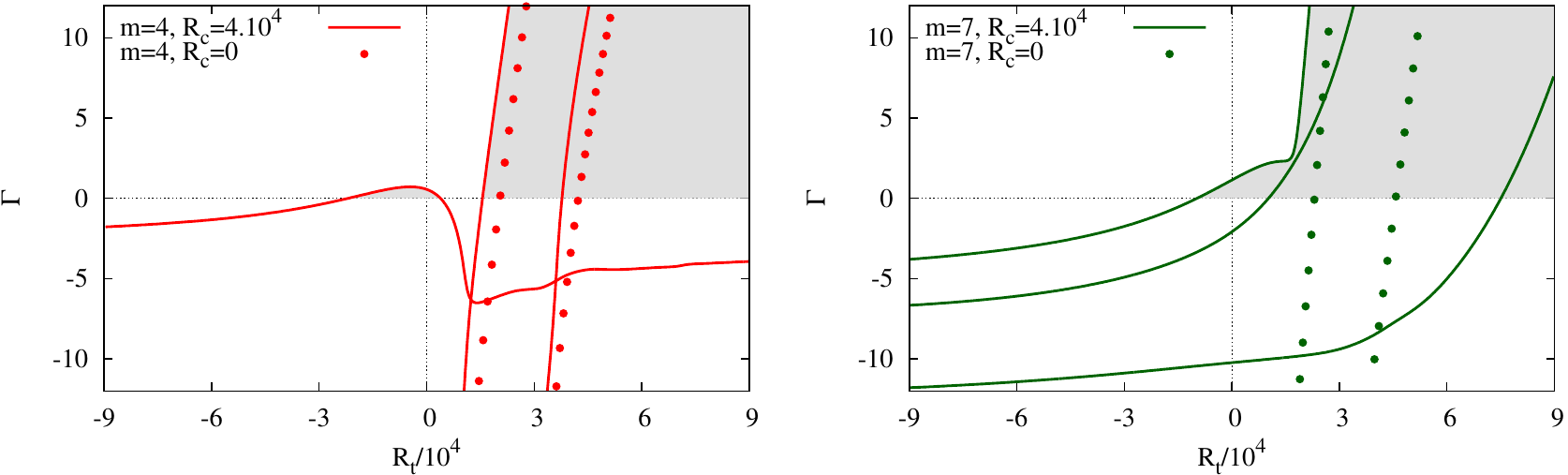}
\caption{{Growth rates of selected eigenmodes of double-diffusive
convection (solid lines) and of a comparable case of purely-thermal
convection (dotted lines) as a function of $\Rt$ for azimuthal wave
numbers $m=4$ in (a) and $m=7$ in (b) and at 
$\tau=1.2\times10^3$, $\Pra=1$, $\Sc=10$, $\eta=0.35$ and at 
$\Rc=4\times10^4$ (double-diffusive case; solid lines) or $\Rc=0$
(purely-thermal case; dotted lines). The red and green colours
correspond colours used in figure~\ref{f:IntroducingRcRtplane} for
$m=4$ and $m=7$, respectively. Convectively unstable regions are shaded.}
(Color online)
} \label{f:eigenModes} 
\end{figure}

The introduction of the equation of concentration in system
\eqref{eq:convection} leads to {the occurrence of additional eigenmodes
not present in the purely thermal case. This is due to} the
increased dimension of the of the eigenvalue
problem~\eqref{eq:algebraic} {compared to the corresponding purely
thermal problem, as is well illustrated in \citep{Simitev2011}}. 
{As an example,} figure~\ref{f:eigenModes} shows a comparison
between {selected} eigenmodes of a {thermo-compositional} case
and the eigenmodes of {the corresponding} purely thermal case for two particular azimuthal wave
numbers $m=4$ and $m=7$ with $\Rc=4\times10^4$, $\tau=1.2\times10^3$ and $\Sc=10$. 
Two types of eigenmodes are observed -- (a) modes that occur in
both the thermo-compositional and the purely thermal case and appear
to correspond to each other, and (b) additional modes that have no
{counterparts} in the purely thermal case. {These latter modes
appear to represent the ``fingering'' and ``diffusive'' instability modes known
from classical studies of double-diffusive convection, see \citep{Radko2013}, but
modified by rotation and spherical geometry, see further below for discussion.}
The purely double-diffusive modes of type (b) can become
unstable in parameter regions disconnected from the main region of
convective instability as indicated by their positive growth rate
shown in the left panel of figure~\ref{f:eigenModes}. 
{This was also observed in the case of the
cylindrical annulus \citep{Simitev2011}.} Whether {such a}
disconnected region will remain present in the global critical curve
depends on the overlap with the instability regions of the eigenmodes
of other wave numbers. This leads us to a  discussion of the
construction of global critical curves below. 

\begin{figure}
\centering
\includegraphics[width=0.7\columnwidth]{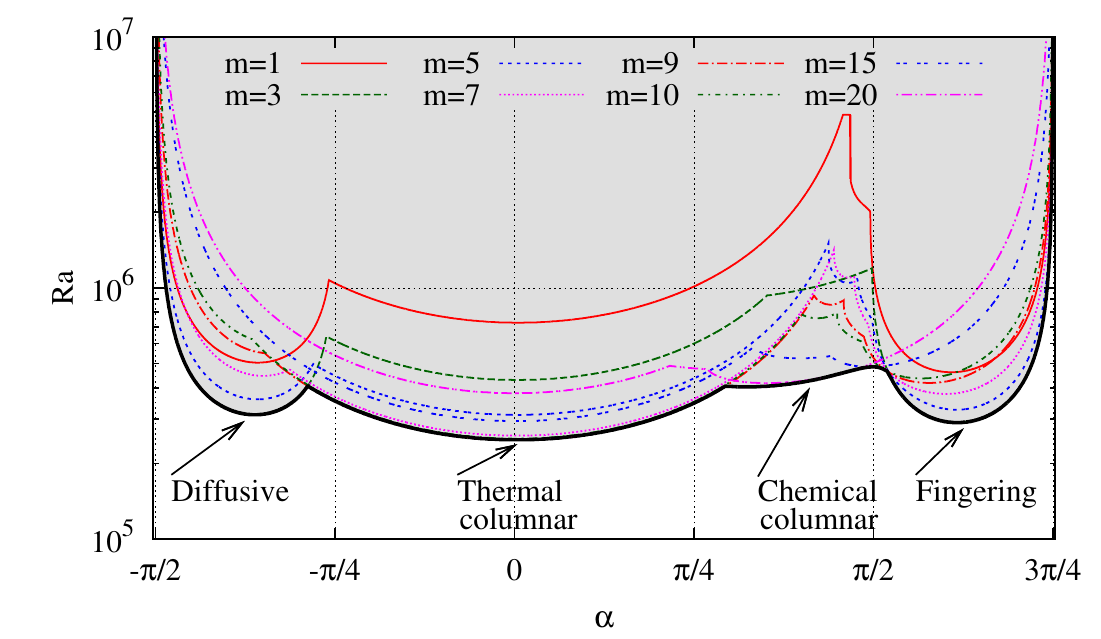}
\caption{Construction of a global critical curve (thick solid black line) as
  the lower envelope of the critical curves for all azimuthal wave numbers $m$. The
  critical curves of individual wave numbers are in turn also
  constructed as the lower envelopes of their structural eigenmodes as
  illustrated in figures~\ref{f:IntroducingRcRtplane} and
  \ref{f:eigenModes}. Results are for $\eta=0.35$, $\tau=10^4$,
  $\Pra=1$ and $\Sc=100$. Only curves for selected values of $m$ are
  shown. The region of convective instability is shaded. {Arrows
    with labels point to distinct branches of the curve identified
    with four convective regimes discussed in the text.}
(Color online)} 
 \label{f:GlobalCriticalCurve}
\end{figure}

{Instability arises whenever an eigenmode attains a positive growth
rate, as illustrated by the shaded regions in figure~\ref{f:eigenModes}.
Thus, for each azimuthal wave-number $m$ a composite critical curve is
obtained which consists of pieces corresponding to the critical curves
of the structural eigenmodes that become unstable
first. Figures~\ref{f:IntroducingRcRtplane} and 
\ref{f:GlobalCriticalCurve} present examples of critical curves for 
fixed values of the wave number $m$. The location of the transition
between instability due to modes of type (a) and instability due
to modes of type (b) is easy to recognise in these figures. 
For instance, one such transition occurs in 
figure~\ref{f:IntroducingRcRtplane}(a) for $m=4$ at
$\Rc=-4.8\times10^4$  where the curve is not smooth. We identify
the arising purely double-diffusive mode as being analogous to a
diffusive instability in a non-rotating layer of cold
fresh water over warm salty water because to the left 
of this transition $\Rc$ is negative and  $\Rt$ is positive which in our model
corresponds to cooler and lighter fluid over warmer and heavier fluid.
A second such transition occurs for the same $m$ at
$\Rc=5.8\times10^4$ where the curve folds back on itself and becomes three-valued. 
We identify the arising purely double-diffusive mode as being analogous to a
fingering instability in a non-rotating layer of warm
salty water over cold fresh water because the substantial part this
branch occupies a region where $\Rc$ is positive and  $\Rt$ is
negative which in our model corresponds to warmer and heavier fluid
over cooler and lighter fluid. A similar fingering branch where the
critical curve folds back on itself is visible in figures 2(a) and
5(a) of \citep{Net2012} in the case of a shell with rigid boundaries
and with differential rather than internal driving, as well as in
figure 4(b) of \citep{Simitev2011} in the rotating annulus
geometry. In the figure of \citet{Simitev2011} the diffusive branch is
also reported. While fingering instability occurs via exchange of
stabilities in the classical doubly-diffusive plane layer model irrespective of
rotation, in the spherical geometry any mode including fingering is
overstable in order to overcome the Taylor-Proundman
constraint. Overstability takes the form of travelling waves similarly
to purely thermal motions \citep{Busse1970}. Figure
\ref{f:IntroducingRcRtplane} also illustrates the appearance of
critical curves when plotted in the $\alpha-\Ra$ plane.}

{Finally, the global critical curve can be conveniently constructed
in the  $\alpha-\Ra$ plane as the lower envelope of the composite
curves for each wave number $m$. In other words, at each value of
$\alpha$ instability occurs at the smallest value of $\Ra$ among the
critical curves for the separate values of $m$. This is illustrated in
the example shown in figure~\ref{f:GlobalCriticalCurve}.} There, we
can also see that the critical curves for different values of $m$
intersect and overlap allowing convection for only certain $m$-values
and not for others. Some azimuthal wave-numbers will contribute to
large portions of the global critical  curve (e.g. $m=9$ in
figure~\ref{f:GlobalCriticalCurve}); others will contribute only
point-wisely (e.g. $m=15$) and others not at all (e.g. $m=1$).  
{Distinct branches of the global critical curve can be easily
  identified and these correspond to the various convective regimes
  discussed above as well as further in the text.}

{In addition to the purely double-diffusive modes of type (b), here
we are also interested in the top-heavy convection modes of type
(a) as these are most likely the modes responsible for the generation
of magnetic field in the core. These modes appear analogous to
purely thermal convection modes as discussed in relation to
figure~\ref{f:eigenModes}.  
To confirm this and to help interpret parameter dependences,} 
{in subsequent sections we compare certain branches of our numerical
solutions to} known {closed-form} {approximations} of
the critical Rayleigh number, azimuthal wave number and drift-rate
{for the onset of columnar convection} in rapidly rotating
{axisymmetric systems}. 
The main difficulties for analytical study of convection in rotating
spheres and shells are the significant variation in the 
Coriolis force  with the angle between gravity and angular velocity
and the inclination of the spherical boundaries with respect to the axis
of rotation. {
In the asymptotic limit $\tau \to \infty$ a so called ``local''
theory of viscous convection in an internally heated sphere was
developed by \citet{Roberts1968} and \citet{Busse1970}.}
\citet{Soward1977} showed that the critical mode predicted by the
local theory is not physically realized as it decays due to
phase mixing. {In particular, in the case of internal heating
  convection onsets deep inside the shell and the true value 
of the critical Rayleigh number may exceed the local theory estimate
by about 25\% \citep{Jones2015}.}  
{Following \citet{Yano1992}, a
``global'' asymptotic theory of the onset of viscous convection was
constructed for rapidly-rotating spheres \citep{JonesSoMu2000} and for spherical shells
\citep{DORMY2004} and results were shown to agree
with accurate numerical solutions of the corresponding linear problems.
Because the global theory does not in general yield expressions in
convenient  closed analytical form, we use here the following less accurate
local theory approximations} 
\begin{align}
\begin{split}
R_{\text{crit}}=7.252\left(\dfrac{P\tau}{1+P}\right)^{4/3} & (1-\eta)^{7/3},\quad
m_{\text{crit}}=0.328\left(\dfrac{P\tau}{1+P}\right)^{1/3}(1-\eta)^{-2/3},
 \\
 \omega_{\text{crit}}&=-0.762\left(\dfrac{\tau^2}{P(1+P)^2}\right)^{1/3}(1-\eta)^{2/3},
\label{highPasym}
\end{split}
\end{align}
where $P$ refers to either the thermal Prandtl number $\Pra$ or the
Schmidt number $\Sc$. {These expressions are derived from equations
(2.7), (4.6) and (4.7) of \citet{Yano1992} by re-scaling in terms of
the dimensionless parameters, length and time scales used in the
present paper. In particular, the factor $(1-\eta) = (r_o-r_i)/r_o =
d/r_o$ enters due to length scale conversion: the radius of the outer
surface $r_o$ is used as length scale by \citet{Yano1992} while the
shell thickness $d=r_o-r_i$ 
is used in our work. In turn, the expressions of \citet{Yano1992}
provide a first-order correction of the local asymptotic results of
\citet{Busse1970} and are thus expected to better account for the
effects of finite inclination of the outer spherical boundary.}
{While expressions \eqref{highPasym} are not valid asymptotic
results for the configuration studied here {we consider them adequate
for the purpose of identifying trends in the numerical results
presented in our paper.} They provide an approximate idea of the
dependence of the critical values for the onset of convection on the
parameters of the problem and similar expressions have been used in
this sense in other studies e.g.~\citep{Simitev2003}.  
}
{In the limit} of $P\to \infty$ (i.e. either $\Pra\to \infty$ or $\Sc\to\infty$) we obtain
\begin{align}
\begin{split} 
\lim_{P\to\infty}R_{\text{crit}}=7.252\,\tau^{4/3}(1&-\eta)^{7/3},\quad
  \lim_{P\to\infty} m_{\text{crit}}=0.328\,\tau^{1/3}(1-\eta)^{-2/3}\\
&\lim_{P\to\infty}\omega_{\text{crit}}\to 0.
\end{split}
\end{align}
The derivatives with respect to $P$ of equation~\eqref{highPasym} are all positive 
\begin{align}
\dfrac{\upartial R_{\text{crit}}}{\upartial P}>0, \quad \dfrac{\upartial m_{\text{crit}}}{\upartial P}>0, \quad \dfrac{\upartial \omega_{\text{crit}}}{\upartial P}>0,
\end{align} 
with all derivatives vanishing as $P\to \infty$. Therefore, we
expect that in the region $0\leq\alpha\leq \pi/2$ where $\Rt$ and
$\Rc$ are positive and specifically at $\alpha=0$ (purely thermal
case) and $\alpha=\pi/2$ (purely compositional case) {the values of
$\Ra$ and $m$ will increase towards  limiting values while $\omega$
will decrease towards zero as $\Pra$ is increased.}
Equations~\eqref{highPasym} also show that
$R_{\text{crit}}\propto \tau^{4/3}$,
$m_{\text{crit}}\propto\tau^{1/3}$ and
$\omega_{\text{crit}}\propto\tau^{2/3}$. Therefore, {we expect that}
the critical Rayleigh and wave numbers and drift-rate will increase as
$\tau$ is increased. Finally, {the $\eta$-dependence in
equations~\eqref{highPasym} is}
$R_{\text{crit}}\propto(1-\eta)^{{7/3}}$,
$m_{\text{crit}}\propto(1-\eta)^{-2/3}$ and
$\omega_{\text{crit}}\propto (1-\eta)^{2/3}$. Therefore, we expect
that the critical Rayleigh number and drift-rate will decrease and
the critical azimuthal wave number will increase
as $\eta$ is increased.
{Equations~\eqref{highPasym} are appropriate only for sufficiently
  large $P\tau\gg1$. At small values of $P\tau$ convection
  takes the form of eddies attached to the outer equatorial boundary
  that can be understood as inertial oscillations modified by
  viscous friction and thermal buoyancy effects. 
Asymptotic expressions for this case can be found in
\citet{Zhang1994,BusseSimitev2004} but will not be used here since
$P\tau$ is kept relatively large.}

\section{Dependence on the Prandtl number}
\label{s:PrDependence}
\subsection{The case of equal Prandtl numbers}
\label{s:samePrandtl}
\begin{figure}
{\footnotesize (a) \hspace{0.48\textwidth} (b)} \\
\subfigure{\resizebox*{0.5\textwidth}{!}%
{\includegraphics[]{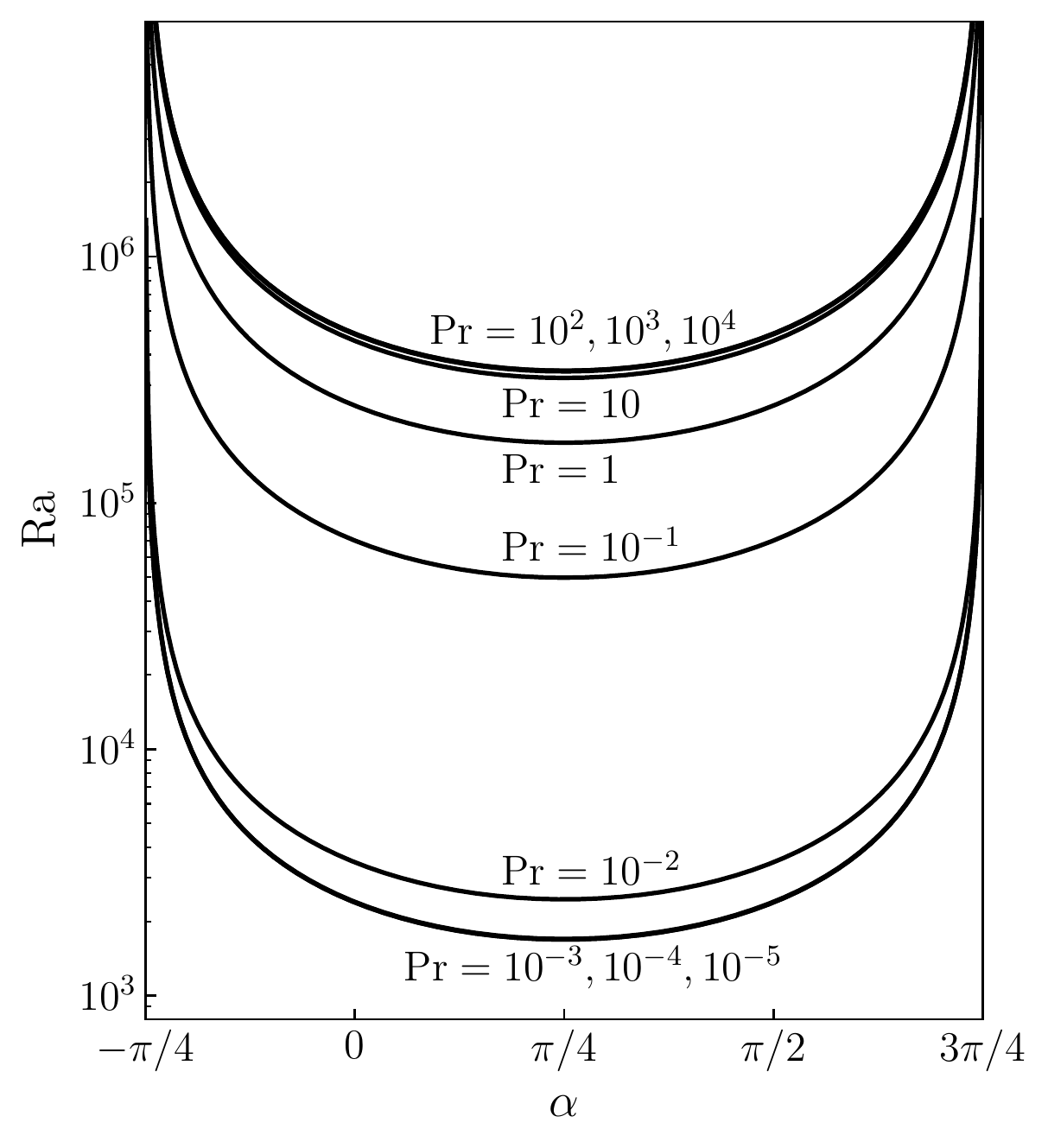}}}
\subfigure{\resizebox*{0.49\textwidth}{!}%
{\includegraphics{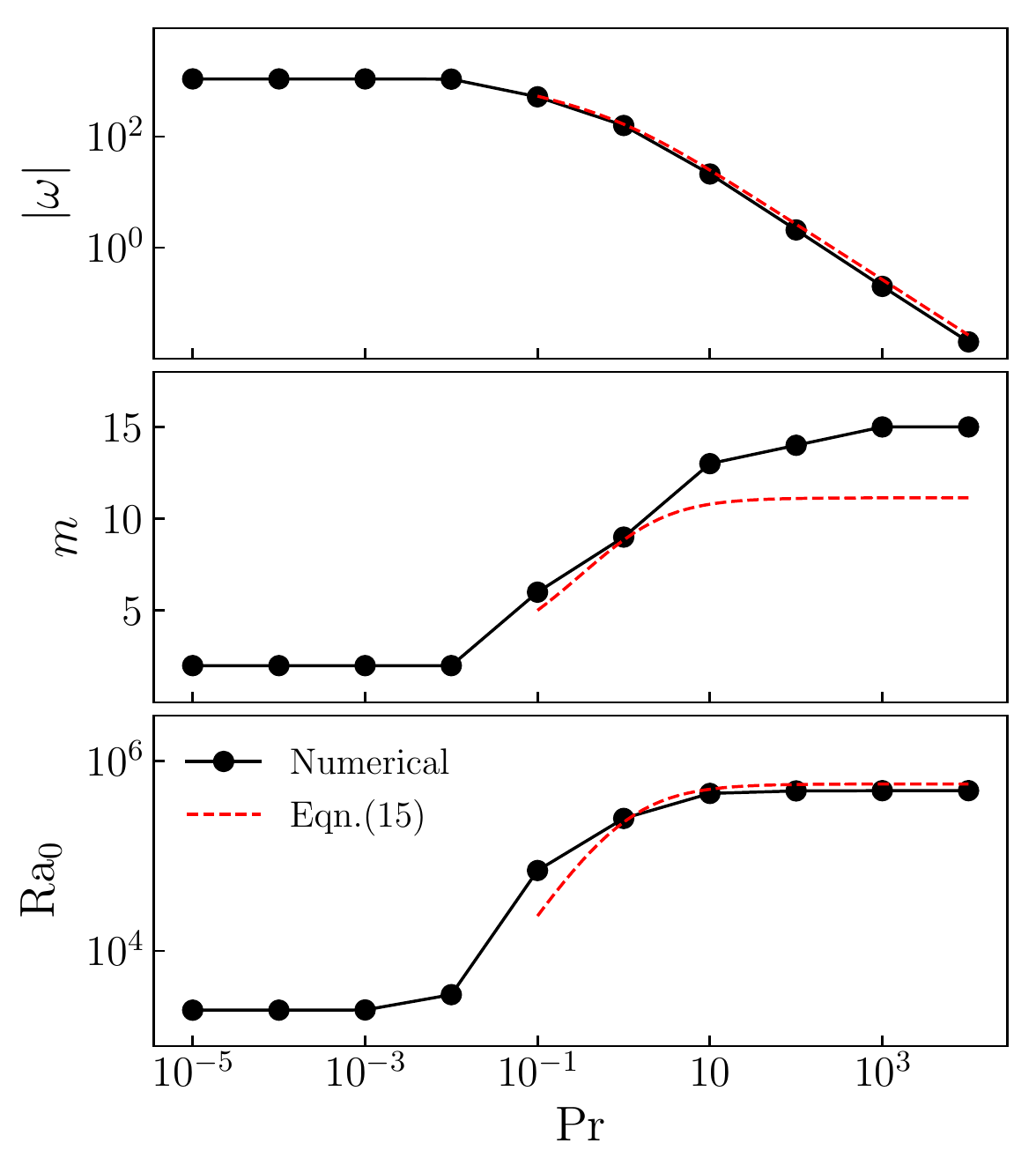}}}
\caption{{Critical parameter values for the onset of convection in the
case $\Pra=\Sc$ and with $\tau=10^4$ and $\eta=0.35$. 
(a) Critical Rayleigh number $\Ra$ as a function of $\alpha$ 
for values of $\Pra$ as indicated in the plot;
Note that the three lowermost curves as well as the three uppermost
curves overlap and appear indistinguishable. (b) Critical
Rayleigh number factor $\Ra_0$ (bottom panel), most unstable
wave-number $m$ (middle panel), and drift-rate amplitude
$|\omega|$ (top panel) as functions of $\Pra$ for
$\alpha=\pi/4$.} {Black solid curves marked with circles are numerical
results and {red dashed curves are values given by the 
approximations \eqref{highPasym}}}. {Circle markers correspond to the values of
the $\Pra$ used in panel (a).}
(Color online)
}
\label{f:dependencePr}
\end{figure}

The simplest case to {consider} is the case of equal Prandtl and Schmidt
numbers as it {is essentially equivalent to the case of} purely
thermal convection. When temperature and composition have the same 
diffusivities the thermal and the compositional Prandtl numbers
{are identical, i.e.~$\Pra = \Sc$, and} composition and
temperature will evolve in a similar manner. 
{In terms of equations~\eqref{eq:newRaFormalism}, the parameter
$\Pminus$ vanishes and as a result} equations
\eqref{eq:newRaFormalismRho} and \eqref{eq:newRaFormalismRhoP}
decouple from each other, the momentum equation
\eqref{eq:newRaFormalismNS} depends only on $\Psi$ while $\Psi'$
becomes a passive tracer for the flow. System
\eqref{eq:newRaFormalism} reduces to the familiar co-density
model. The critical Rayleigh number dependence {takes the form 
of a} straight line $\Rt=-\Rc+\Ra_0$ in the $\Rt-\Rc$ plane, as
proposed by \citet{BreuerEtAl2010}, and in the $\Ra-\alpha$ plane we obtain
\begin{equation}
 \Ra_c = \frac{\Ra_0}{\cos\alpha + \sin\alpha}.
 \label{eq:RaBreuer}
\end{equation}
Figure~\ref{f:dependencePr}(a) shows the critical curves $\Ra_c$ as a
function of $\alpha$ for $\Pra$ varying between $\Pra=10^{-5}$  (lower
curves) to $\Pra=10^4$ (higher curves) {and for values of the
Coriolis number and the shell aspect ratio fixed to $\tau=10^4$ and
$\eta=0.35$ (the dependence on $\tau$ and $\eta$ is discussed in later
sections).} All curves are symmetric with respect to $\alpha=\pi/4$
(equal Rayleigh numbers) and {approach} infinity with asymptotes
at $\alpha=-\pi/4$ and $\alpha=3\pi/4$. 
Equations~\eqref{eq:newRaFormalismNS} and \eqref{eq:newRaFormalismRho}
show that the  larger $\Pra$ becomes, the smaller its influence
is. The coefficient $\Ra_0$ in \eqref{eq:RaBreuer} should then become
independent of $\Pra$ at large values of this parameter. We will
{refer to this regime   as} the advective regime. At Prandtl
numbers much smaller than one the process of diffusion takes over and
prevents small scale convection from growing. Then much  larger scales
dominate and  the onset of convection occurs at lower values of the
Rayleigh numbers. {We will refer to this regime as} the diffusive
regime.  {It is easy to recognise that the advective and the
diffusive regimes are, in fact, the familiar columnar convection
regime \citep{Busse1970} and inertial convection \citep{ZhangBusse87,Zhang1994} regime, respectively.} 
The transition between these two regimes occurs at {about
  $\Pra=0.1$ for $\tau=10^4$} \citep{ArdesEtAl97,BusseSimitev2004}.   
{To confirm this} we plot the critical values of $\Ra_0$, of the
azimuthal wave number $m$ and of the drift rate $\omega$ as a function
of $\Pra$ in figure~\ref{f:dependencePr}(b). The values of $\Ra_0$ and
the $m$ saturate in the limits $\Pr\to 0$ and $\Pr\to\infty$, with
small {values of} $m$ ($m=2$) for $\Pr\to 0$ and large {values of} $m$
($m\geq14$) for $\Pr\to\infty$. The transition between the diffusive and
the advective regime is easy to identify in
figure~\ref{f:dependencePr}(b). We have also plotted for comparison
the approximations~\eqref{highPasym} for the critical
parameter of columnar convection. As noted previously, 
approximations for the critical parameters of inertial convection are
also available \citep{Zhang1994,BusseSimitev2004} {but have not be
included in figure~\ref{f:dependencePr}(b)}.

{The critical modes exhibit the typical spacial features 
of inertial convection for smaller values of $\Pr$ and of
columnar convection for larger values of $\Pr$.} For lower values of
the $\Pr$, convection takes the form of large non-spiralling convection eddies
attached to the outer boundary of the shell near the equator and exhibit a strong
clockwise drift.
Above the abrupt transition at about $\Pra=0.1$ convection columns
with much smaller azimuthal scale appear. {Near the transition the
structure of columns resembles} the double-humped mode 
described by \citet{ArdesEtAl97}. By $\Pra=0.2$ columnar convection is
fully established and show a typical spiralling
shape. {Convection columns are detached from outher spherical
surface and are arrange in a cartridge belt immediately outside of
the tangent cylinder.} {Inertial and columnar convection regimes
with properties similar to these described here for $\tau=10^4$  exist also for more general values
of $\tau$ \citep{Zhang1994,Simitev2003}.}

\subsection{Small departures from the $\Pra=\Sc$ regime}
\label{s:smallDepartures}
\begin{figure}
{\footnotesize (a) \hspace{0.49\textwidth} (b)} \\
\subfigure{\resizebox*{0.495\textwidth}{!}%
 {\includegraphics{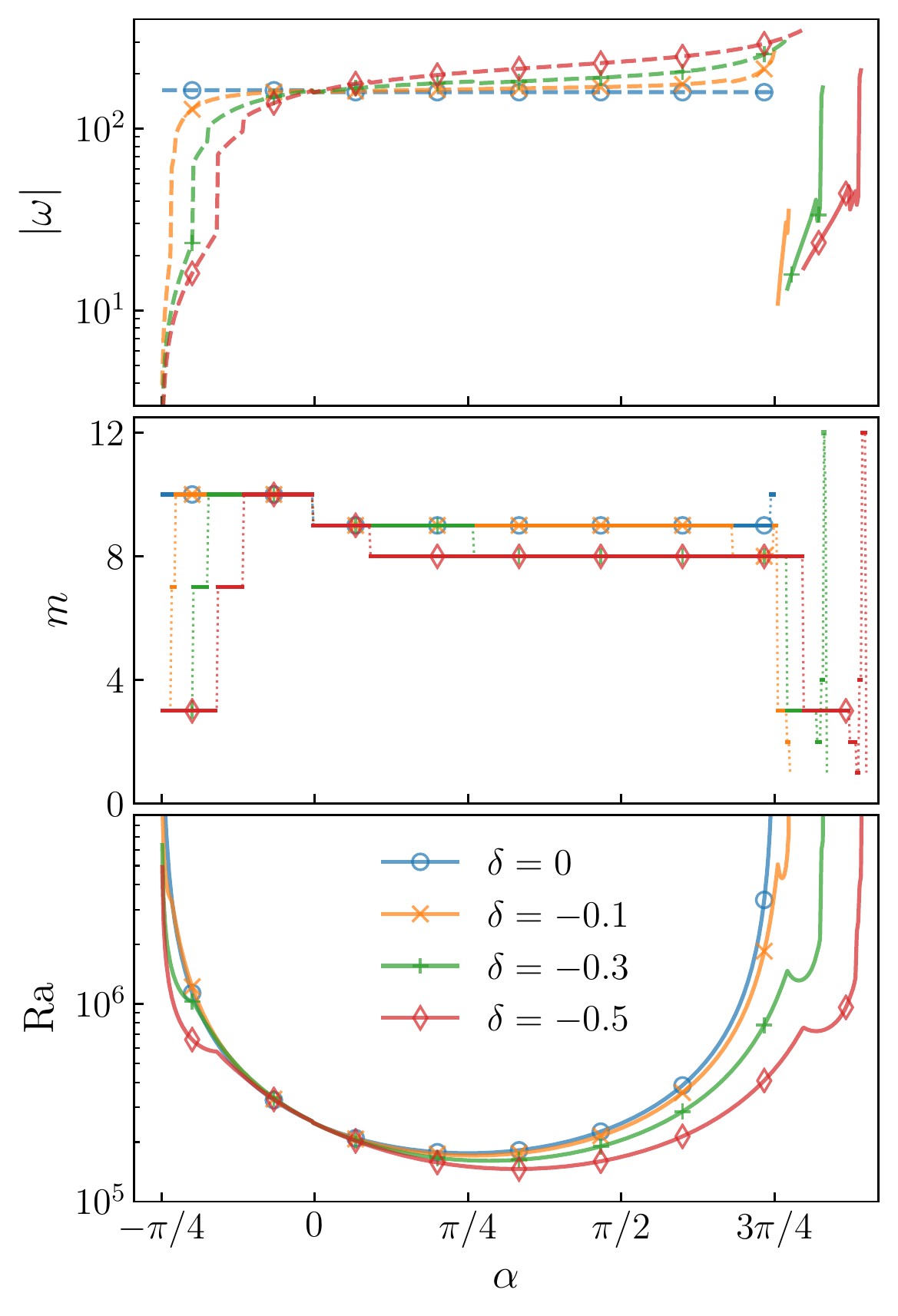}}}
\subfigure{\resizebox*{0.505\textwidth}{!}%
{\includegraphics{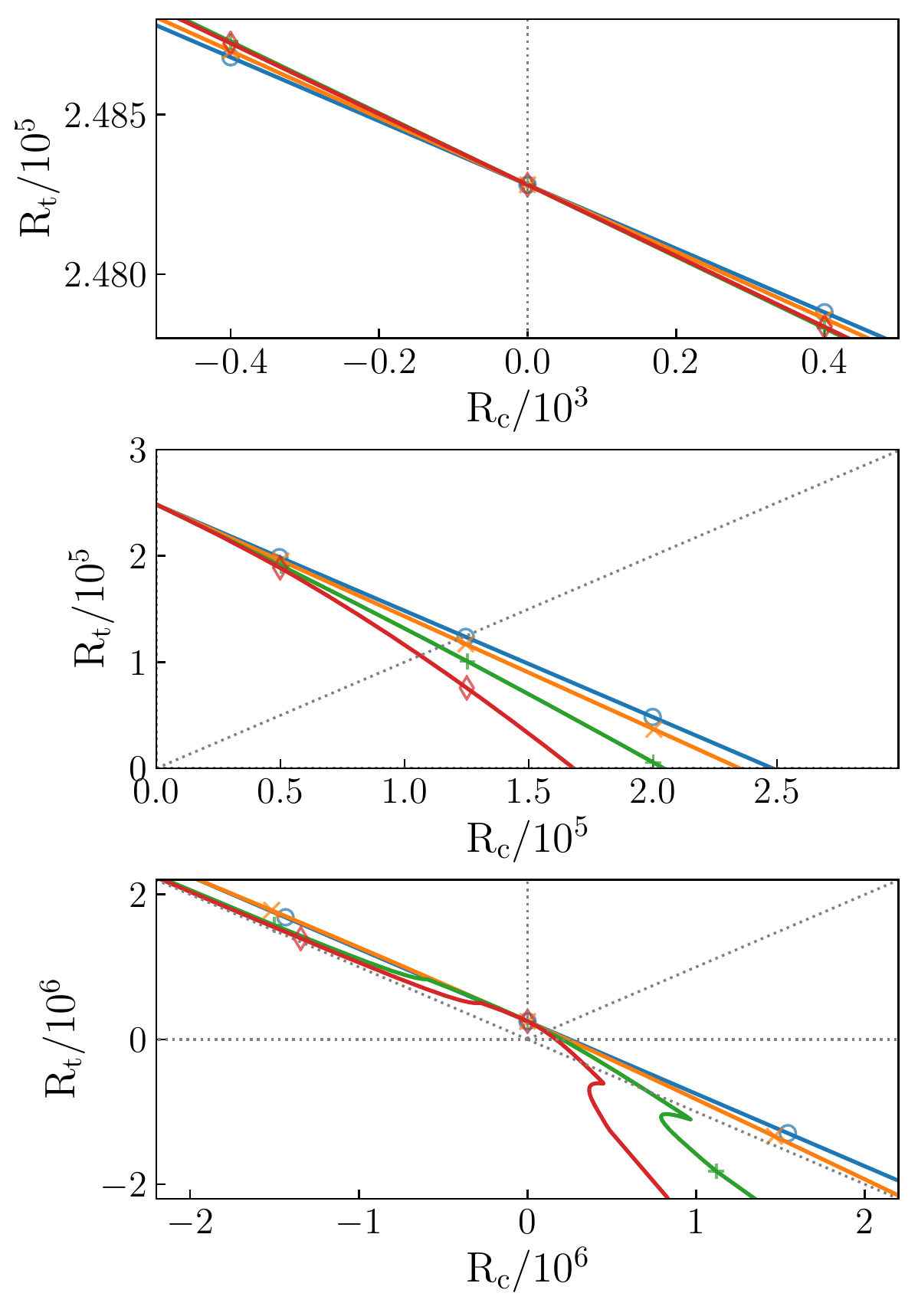}}}
\caption{ 
{Critical parameter values for the onset of convection at $\tau=10^4$,
$\eta=0.35$, $\Pra=1$ and $\Sc = \Pra(1+\delta)$ for several values of
$\delta$ as specified in the legend. (a) Rayleigh number $\Ra$ (bottom
panel), most unstable wave-number $m$ (middle panel), and drift-rate
amplitude $|\omega|$ (top panel) as a function of $\alpha$. 
Negative values of $\omega$ are indicated by dashed lines and positive
values are indicated by solid lines. 
(b) Critical curves in three representative regions of the \Rc-\Rt
plane. Thin grey dotted lines included for reference correspond to
$\alpha=k\pi/4$, $k=-2\dots3$ in anticlockwise direction (lowermost panel).}
(Color online)
}
\label{f:Le.ap.1}
\end{figure}

\begin{figure}
{\footnotesize (a) \hspace{0.49\textwidth} (b)} \\
\subfigure{\resizebox*{0.497\textwidth}{!}
 {\includegraphics{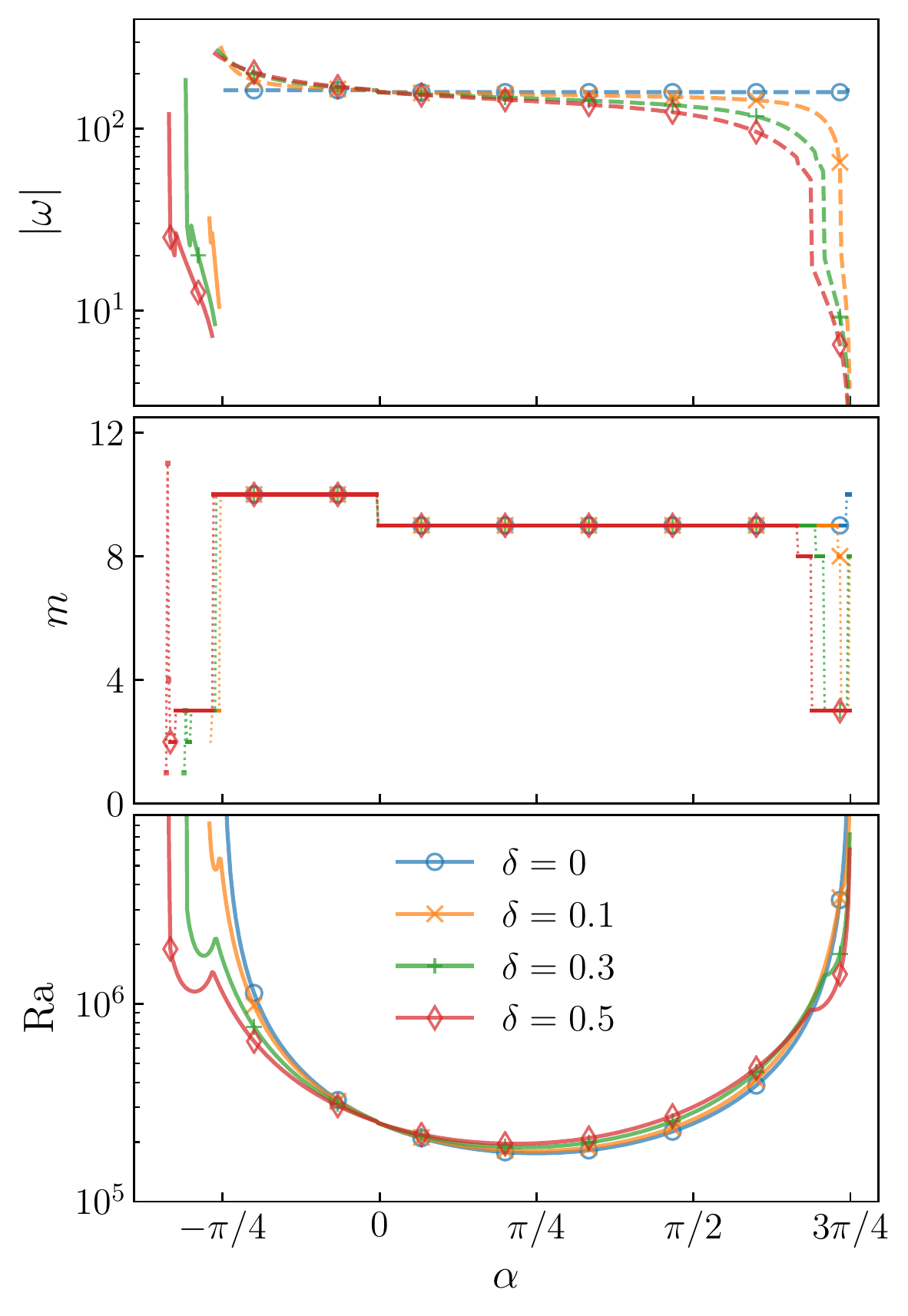}}}
\subfigure{\resizebox*{0.503\textwidth}{!}
{\includegraphics{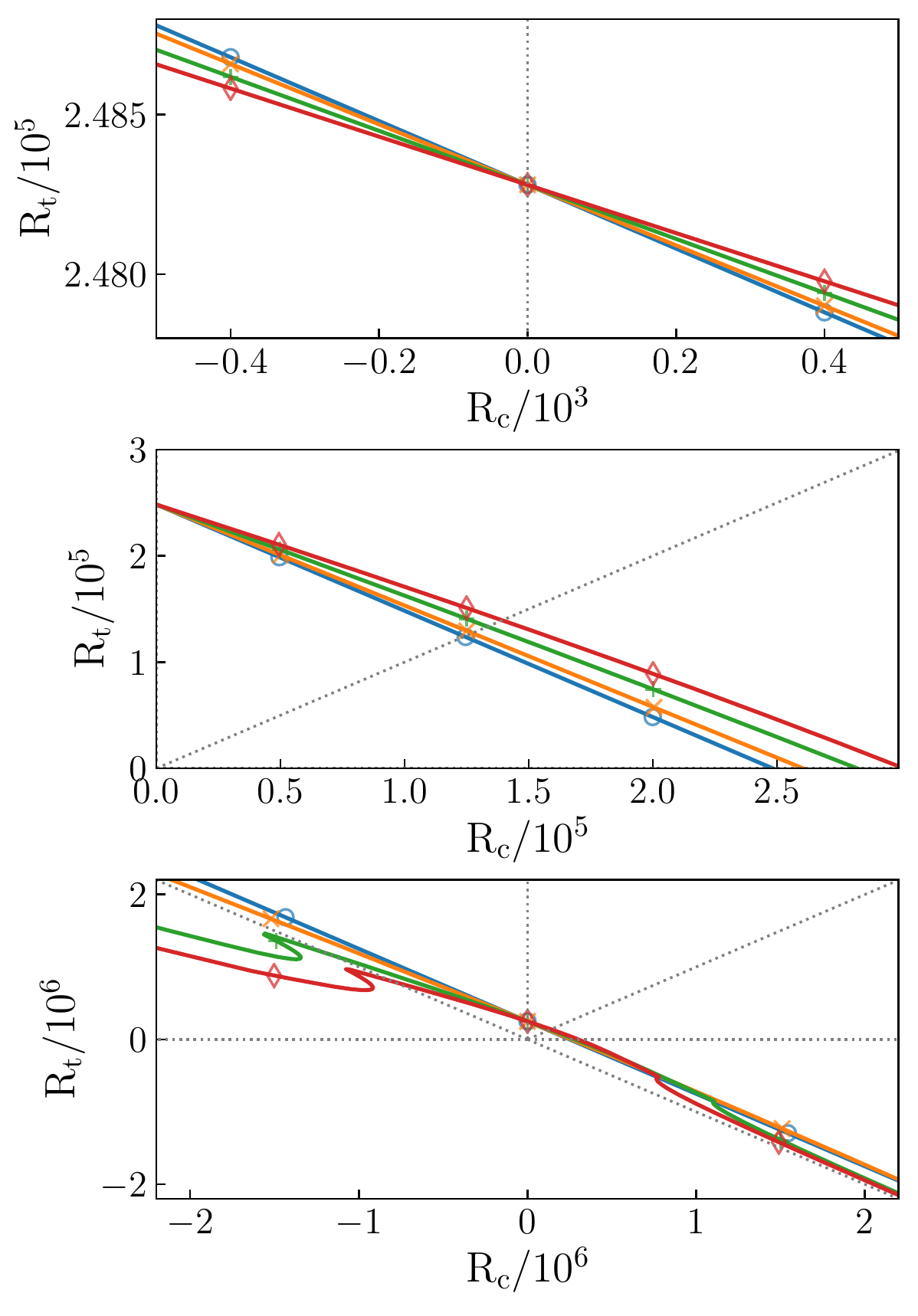}}}
	\caption{Same as figure~\ref{f:Le.ap.1} but for four positive
          values of $\delta$ as specified in the legend. (Color online)}
	\label{f:Le.ap.1posd}
\end{figure}

We now depart from the co-density case of a ``single-diffusive'' fluid
characterised by equal Prandtl numbers. 
In this section we develop a first order approximation to
equations~\eqref{eq:newRaFormalism} that will allow for the analysis 
of small departures from the regime of $\Pra=\Sc$. {We consider a small
number $|\delta|\ll1$} such that the Prandtl and Schmidt numbers are
similar in value 
\begin{gather}
\label{eq:delta}
\Sc = \Pra(1+\delta).
\end{gather}
{To first order in $\delta$} the modified Prandtl numbers
introduced in equations \eqref{eq:NewPrandtls} become
\begin{gather}
 \Pplus^{-1} \approx \frac{\Pra^{-1}}{2} \left(2-\delta\right),
 \qquad
 \Pminus^{-1} \approx \frac{\Pra^{-1}}{2} \delta,
\end{gather}
{with} $\Pplus^{-1}$ being always much larger than $\Pminus^{-1}$ but the 
latter still finite. {With these approximations}
equations~\eqref{eq:newRaFormalismRho} and \eqref{eq:newRaFormalismRhoP} become to $O(\delta)$,
\begin{subequations}
\label{eqpnumb}
\begin{align}
	\upartial_t\Psi &= \Pra^{-1}\nabla^2\Psi-{\bm u}{\bm{\cdot}}{\bm{\nabla}}\Xi+\frac{\delta}{2}\Pra^{-1}\nabla^2(\Psi'-\Psi),\label{eqpnumb1}\\
		\upartial_t\Psi' &= \Pra^{-1}\nabla^2\Psi'-{\bm u}{\bm{\cdot}}{\bm{\nabla}}\Xi'+\frac{\delta}{2}\Pra^{-1}\nabla^2(\Psi-\Psi').\label{eqpnumb2}
\end{align}
\end{subequations}
{Using definitions \eqref{eq:newRaFormalismVars}} and the reference
temperature and composition profiles equations \eqref{eqpnumb} can
be rewritten in the form
\begin{subequations}
\label{eqpnumb:22}
\begin{align}
	\upartial_t \Psi &=\Pra^{-1}\nabla^2\Psi+\beta u_r(\cos \alpha + \sin \alpha)-\delta\sin\alpha\nabla^2 \chi, \label{eqpnumb3}\\
	\upartial_t\chi  &=\Pra^{-1}\nabla^2\chi+\beta u_r-\delta \Pra^{-1}\nabla^2\chi,\label{compappeq}
\end{align}
\end{subequations}
{where} $\beta=\beta_{\mathrm{T}}=\beta_{\mathrm{C}}$ was assumed.
{Equations~\eqref{eqpnumb:22}} are only coupled to each other by
the last term in ~\eqref{eqpnumb3} which is multiplied by the small
quantity $\delta$ and so the diffusion of $\chi$ has only a small
contribution to the evolution of the buoyancy profile $\Psi$.  
It is now easy to understand the physical effects of deviating from
the case of equal Prandtl numbers as discussed bellow. {These are
illustrated in} figures~\ref{f:Le.ap.1} and \ref{f:Le.ap.1posd} for
values of the Schmidt number $\Sc$ just below and just above $\Pra=1$,
respectively, and for fixed values of the Coriolis number and the
shell aspect ratio $\tau=10^4$ and $\eta=0.35$.

When $\Sc<\Pra$, so that $\delta<0$, and $0<\alpha<\pi/2$ (positive
compositional and thermal Rayleigh numbers) equation~\eqref{eqpnumb3}
shows that the growth in time of the compositional component $\chi$
has a small but destabilising effect on the buoyancy profile $\Psi$.
Thus convection can now occur for lower values of $\Ra$ as $|\delta|$ 
becomes larger. This is evident in the bottom panel of
figure~\ref{f:Le.ap.1}(a) where, upon decreasing delta from 0 to
$-0.5$, a reduction in $\Ra$ is indeed observed. The magnitude of this  
reduction increases as  $\alpha$ is increased from 0 to
$\pi/2$. The middle panel of figure~\ref{f:Le.ap.1}(b) {shows this
effect in the $\Rt-\Rc$ plane.} More negative values of $\delta$
correspond to smaller values of $\Sc$ so that less energy is required
for the onset of convection and both critical Rayleigh numbers
{(thermal and compositional)} decrease. This confirms the results of
\citet{Simitev2011} where for positive Rayleigh numbers, an increase
in $\Rc$ leads to a reduction in the critical value of $\Rt$ required
for the onset of convection.  

When $\Pra<\Sc$, so that $\delta>0$, and $0<\alpha<\pi/2$ the
diffusion of $\chi$ has a stabilising effect on the buoyancy profile,
$\Psi$, with convection occurring at higher critical values of $\Ra$
as $\delta$ increases. This effect is {evident} in the bottom
panel of figure~\ref{f:Le.ap.1posd}(a), where an increase in $\Ra$ is
seen as $\delta$ is increased. 
Similarly, this effect becomes larger as $\alpha$ is increased from 0 to 
$\pi/2$ as illustrated in the lower panel of
figure~\ref{f:Le.ap.1posd}(a) and in the middle panel of
figure~\ref{f:Le.ap.1posd}(b). 

{For sufficiently small negative values of alpha}, $\alpha<0$,
$|\alpha|\ll1$ and $\Sc<\Pra$ the sign of $\sin \alpha$ is negative
and the diffusion of $\chi$ has a stabilising effect as $\delta \sin\alpha$ is positive. This is illustrated in the
top panel of figure~\ref{f:Le.ap.1}(b) where as $\alpha$ changes from
positive to negative and we see that decreasing $\delta$ increases the
critical value of the thermal Rayleigh number $\Rt$ whereas when
$\alpha>0$ decreasing $\delta$ decreases the critical
$\Rt$. Similarly, when $\alpha>0$, $|\alpha|\ll1$ and $\Sc>\Pra$ the
diffusion of $\chi$ leads to a reduction of the buoyancy profile. This
is seen most clearly in the upper panel of figure~\ref{f:Le.ap.1posd}(b). Increasing $\delta$ leads to a reduction of
the critical value of $\Rt$. This phenomenon can be explained by the
fact that, when $\alpha<0$, the compositional buoyancy is now acting
against the thermal component of buoyancy. Therefore, an increase in
$\Sc$ inhibits the effect of the compositional component and thus
lowers the critical value of $\Rt$.     

{In summary,} the departure from equal Prandtl numbers has a small
but noticeable affect on the onset of convection, manifested by a
deviation from the ``co-density'' values of $\Ra$  obtained in the
case $\Pra=\Sc$. However, the regions where $\Rt$ and $\Rc$ have
opposite signs, namely $-\pi/2 < \alpha < 0$ and $\pi/2 < \alpha <
\pi$, will be most strongly affected as a delicate balance exists
there between stabilising and destabilising forces.   

The lower panels of figures~\ref{f:Le.ap.1}(a) and
\ref{f:Le.ap.1posd}(a) {show} that the asymptotic behaviour 
{exhibited by} the critical curves $\Pra=\Sc$ as $\alpha
\rightarrow-\pi/4$ and $\alpha \rightarrow3\pi/4$ is followed on only
one side by the critical curves with $\Pra\ne\Sc$. {When $\Sc>\Pra$} the
asymptote at $\alpha \rightarrow-\pi/4$  is approached, when
$\Sc<\Pra$ the asymptote at $\alpha \rightarrow3\pi/4$ asymptote is approached.
{Since} the shift of the asymptote is due to coupling
between equations~\eqref{eqpnumb:22},
we consider small departures from the values $\alpha=-\pi/4$ and $3\pi/4$ by setting
$\alpha= -\pi/4+\epsilon$ and $3\pi/4 +\epsilon$, where
$|\epsilon|\ll1$.  Expanding $\cos\alpha$ and $\sin\alpha$ in Taylor series around these
values equation~\eqref{eqpnumb3} becomes
\begin{align}
\upartial_t\Psi={\Pra}^{-1}\nabla^2\Psi\pm\epsilon\sqrt{2}\beta u_r\pm\delta\dfrac{\Pra^{-1}}{\sqrt{2}}\nabla^2\chi.\label{-pi/4}
\end{align}
where, the upper and lower signs of the $\pm$ term refers to $\alpha=-\pi/4$ and $3\pi/4$, respectively.
We now consider the two cases $\Pra>\Sc$ ($\delta<0$) and $\Pra<\Sc$
($\delta>0$) in turn. 
Firstly, when $\Pra>\Sc$ and $\alpha=3\pi/4+\epsilon$ with $\epsilon>0$ (the
negative of $\pm$ is taken) the diffusion of $\chi$ is destabilising and
the advective term $\epsilon\sqrt{2}\beta u_r$ is stabilising. Clearly
when $\delta=\epsilon=0$ the buoyancy profile {simply} diffuses away. As
$\delta$ is decreased the {gradient} of the buoyancy profile is
increased and thus $\epsilon$ can take greater values until the advective term
becomes greater than the diffusion of $\chi$. This is supported by the
bottom panel of figure~\ref{f:Le.ap.1}(a) where the asymptote of $\Ra$ is
shifted further to the right as $|\delta|$ increases. This
behaviour is also observed in the bottom panel of figure~\ref{f:Le.ap.1}(b) {where a shift is seen} away from the line
$\alpha=3\pi/4$ towards $\alpha=\pi$ as $\delta$ decreases from $0$ to
$-0.5$ and thus lower critical values of $\Rc$ are required. 
Secondly, when $\Pra<\Sc$ the asymptote shift is mirror reflected at
$\alpha=\pi/4$ and occurs to the left near $\alpha=-\pi/4$. When $\delta>0$ and
$\alpha=-\pi/4+\epsilon$ {with   $\epsilon<0$}, the
positive sign of equation~\eqref{-pi/4} shows again that the diffusion
of $\chi$ and the advective term are destabilising and stabilising,
respectively. Clearly, when the diffusive term can overcome the
advective term, convection can occur. Therefore, as $\delta$ is
increased the asymptote will shift, with negative values of $\epsilon$
of higher magnitude able to still support convection. This continues
with increasing $\epsilon$ until the magnitude of the advective term
becomes too large and convection is unable to occur. The bottom panels
of figures~\ref{f:Le.ap.1posd}(a) and figures~\ref{f:Le.ap.1posd}(b)
support this and show a shift of the asymptote from $\alpha=-\pi/4$ to 
$\alpha=-\pi/2$ and thus lower critical values of $\Rt$ are required. 

An interesting feature is the jump from small scale to
large scale convection immediately beyond the co-density asymptotes
($\alpha=-\pi/4, 3\pi/4$) of the case $\Pra=\Sc$. The azimuthal
wave-number decreases significantly beyond the co-density asymptote
for non-zero $\delta$ as seen in the middle panels of
figures~\ref{f:Le.ap.1}{a} and \ref{f:Le.ap.1posd}{a} and a large drop in
magnitude of the drift-rate $\omega$ is associated with this (see the
top panels of figures~\ref{f:Le.ap.1}a and
\ref{f:Le.ap.1posd}a). However the wave-number and drift-rate quickly
rise towards the new asymptotes. 
Another interesting effect occurs on the opposite side to the
asymptote shifts. Here the co-density asymptotes are still in effect
($\delta=0$) but as $|\delta|$ is increased there is a reduction in
the critical Rayleigh number. Again we the switch from small scale to
large scale convection around these asymptotes. 

Finally, we mention that the calculations shown in
figures~\ref{f:Le.ap.1} and \ref{f:Le.ap.1posd} are for the specific
value $\Pra=1$ but our simulations show that similar results are
obtained for other values of $\Pra$. 

\subsection{Large differences between Prandtl numbers}
\label{s:largePrDifferences}
{In this section we investigate the case} when the values of the
Prandtl and the Schmidt numbers, $\Pra$ and $\Sc$, are significantly
different from each other. 
First, recall that equations~\eqref{eq:PrWhenPt.ne.Pc} and
\eqref{eq:rhoWhenPt.nePctot} {model} the evolution of the buoyancy
field {in this case.} 
For brevity, we will only analyse the case
of $\Sc\gg\Pra$ as the case of {$\Sc\ll\Pra$} is symmetric  with
respect to $\alpha=\pi/4$. {In particular, we fix the values of
the Schmidt number to $\Sc = 25 \Pra$, the Corriolis number to
$\tau=10^4$ and the  shell aspect ratio to $\eta=0.35$ in this section.}
Here equation~\eqref{eq:rhoWhenPt.ne.Pc} becomes 
\begin{equation}
\upartial_t \Psi\approx \dfrac{\Pra^{-1}}{2}\nabla^2(\Psi+\Psi')-{\bm u}{\bm{\cdot}}{\bm{\nabla}}\Xi.
\end{equation}
The evolution of $\Psi$ is now essentially only affected by the diffusion of the
physical field which is characterised by the smaller of the values of
$\Pra$ and $\Sc$. {Since} convection regimes are {rather}
different for values of the Prandtl numbers above and below 0.1 we 
present our analysis in these two cases. {At Prandtl numbers
smaller than 0.1} convection is generally large scale and three
distinct regions of $\Ra_c(\alpha)$ can be identified. {At Prandtl
  numbers greater than 0.1} instead of three, four distinct regions 
{emerge} all which are functions of $\alpha$. Figure~\ref{f:Pc.ggt.Pt}
shows the  results of our 
calculations. 

\subsubsection{Small Prandtl number case}
{At} the lowest value of $\Pra$ in figure~\ref{f:Pc.ggt.Pt}(a),
$\Pra=10^{-5}$, the critical curve $\Ra$ is approximately symmetric
with respect to $\alpha=\pi/4$ for values of the mixing angle from
$\alpha=-\pi/4$ to $\alpha=3\pi/4$. 
{In this range} the null curve is straight line of negative
gradient {when plotted} in the $\Rc-\Rt$ plane. At values $\Pra=10^{-5}$
and $\Sc=2.5\times10^{-4}$ there is little difference in how
thermal and compositional components affect convection which is large scale 
($m=2$) with a {prograde} drift ($\omega<0$) and is of equatorially
attached type. Figure~\ref{f:Pc.ggt.Pt.flowStruct} shows the stream
function ($1/r \upartial {\cal S}/\upartial\theta$) in the equatorial plane for
the case of $\Pra=10^{-5}$ (top left). This situation is easy to
understand in terms of equation~\eqref{eq:newRaFormalismRho} as any
small scale buoyancy anomalies are  rapidly diffused away due to small
values of the Prandtl numbers; the  advection of the background
profiles  cannot overcome such strong diffusion. 

The behaviour in the region between mixing angles {$-\pi/2<\alpha$}  and
 $\alpha<-\pi/4$ is also easy to interpret -- here  
the buoyancy profile $\Xi$ has a negative sign ($\cos\alpha+\sin\alpha<0$) and thus 
acts to stabilise the system. In fact, convection is only possible 
because the advection profile $\Xi'$ is now approaching a maximum and therefore 
generates enough $\Psi'$ to fuel the production of $\Psi$ in 
equation~\eqref{eq:newRaFormalismRho} and consequently the velocity ${\bm{u}}$ in 
equation~\eqref{eq:newRaFormalismNS}. This, however, comes at the cost
of requiring a higher value of $\Ra$ for the onset of convection to
take place. 

\subsubsection{Intermediate Prandtl number case}
\begin{figure}
{\footnotesize (a) \hspace{0.5\textwidth} (b)} \\
\subfigure{\resizebox*{0.481\textwidth}{!}%
 {\includegraphics{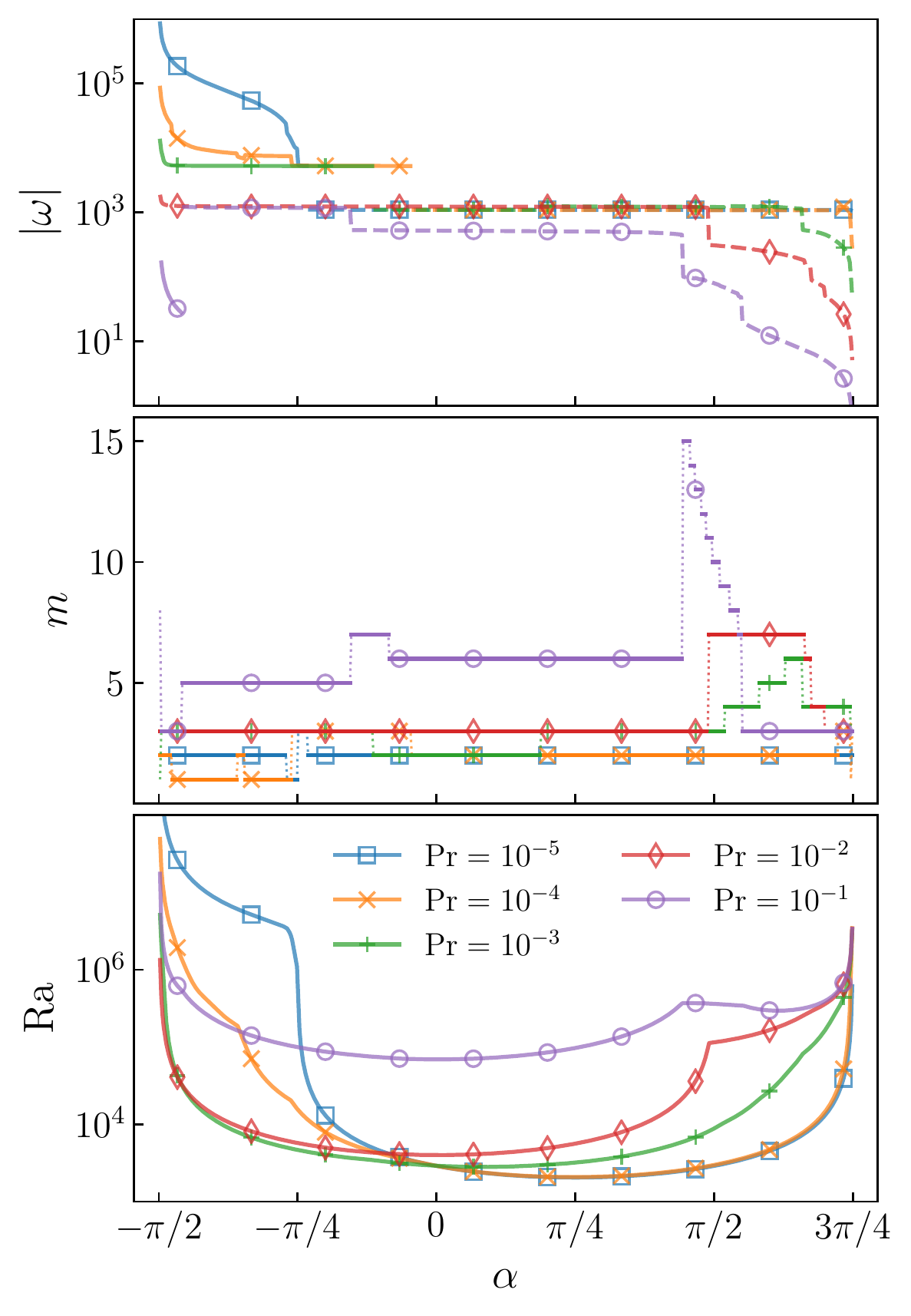}}}
\subfigure{\resizebox*{0.49\textwidth}{!}%
{\includegraphics{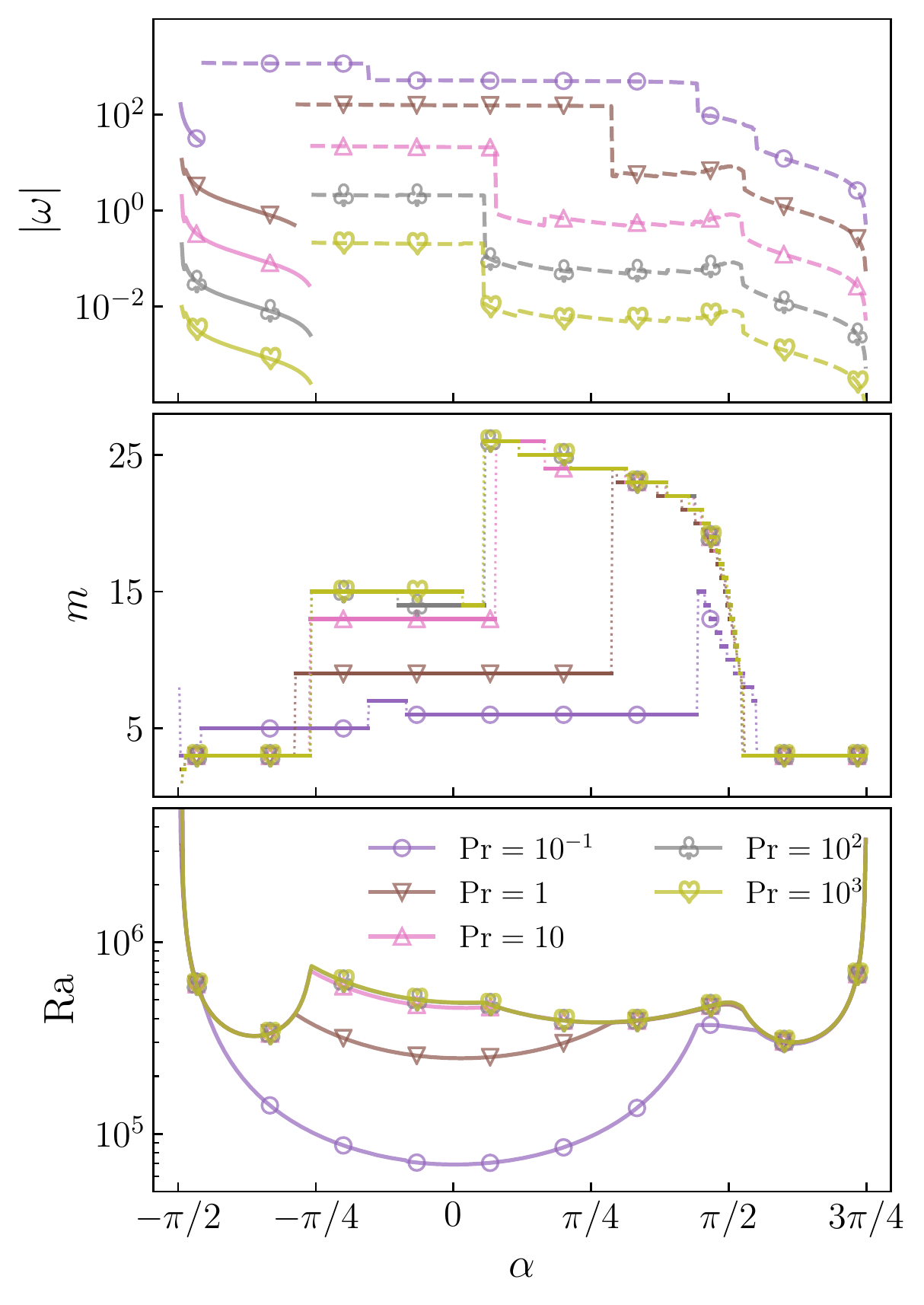}}}
\caption{
{Critical parameter values for the onset of convection at $\tau=10^4$,
$\eta=0.35$, $\Sc = 25\Pra$ and values of the Prandtl number $\Pra$ as
specified in the legends. (a and b) Rayleigh number $\Ra$ (bottom
panel), most unstable wave-number $m$ (middle panel), and drift-rate
amplitude $|\omega|$ (top panel) as a function of $\alpha$. 
Negative values of $\omega$ are indicated by dashed lines and positive
values are indicated by solid lines. 
}
(Color online)
}
 \label{f:Pc.ggt.Pt}
\end{figure}

Increasing $\Pra$ from $10^{-5}$ to $10^{-2}$ in
figure~\ref{f:Pc.ggt.Pt}(a) results in a general shift of the critical
$\Ra$ curve upwards and to the left, {as most clearly seen by
comparing the curves $\Pra=10^{-5}$ and $\Pra=10^{-2}$}. 
Wave number values increase gently from $m=3$ to $m=7$ so that convection is
now on a slightly smaller scale.  
{At} $\Pra=10^{-1}$ {an overall increase
in the critical value of $\Ra$ occurs due to the transition from
inertial convection to columnar convection.}
An interesting effect is that, as we switch from thermally controlled
to compositionally controlled convection at about $\alpha=\pi/3$, there
is a {transition} from large scale ($m=6$) to a {significantly
smaller} scale mixed convection ($m=15$) which has already been
observed by \cite{TrumperEtAl2012} with $\mathrm{Le}=30$. As $\alpha$
increases towards $\pi/2$ (purely compositional convection),
the value of $m$ decreases again to $m=10$ in analogy to
purely thermal convection with internally distributed heat sources and
a Prandtl number of 2.5.  

\begin{figure}
\centering
\begin{tabular}{c@{\extracolsep{15mm}}c}
\begin{overpic}[height=0.24\columnwidth,tics=10]{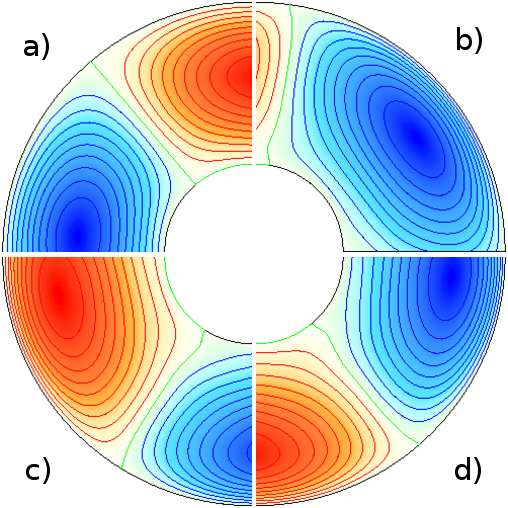}
\put (35,105) {\footnotesize{$\Pra=10^{-5}$}}
\put (-15,70) {\footnotesize{$\displaystyle-\frac{3\pi}{8}$}}
\put (100,70)  {\footnotesize{$\displaystyle 0$}}
\put (-6,25) {\footnotesize{$\displaystyle\frac{\pi}{2}$}}
\put (98,25)  {\footnotesize{$\displaystyle\frac{5\pi}{8}$}}
\end{overpic} &
\begin{overpic}[height=0.24\columnwidth,tics=10]{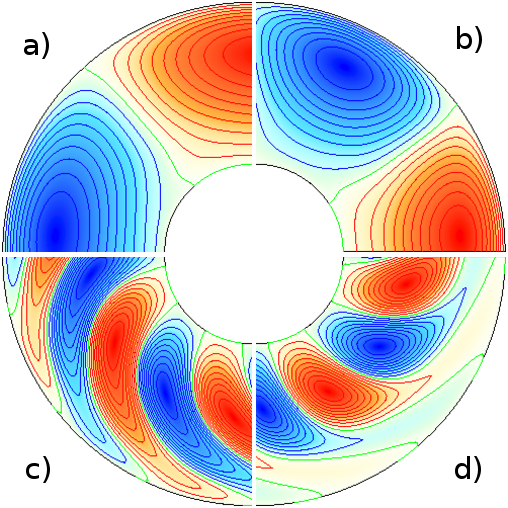}
\put (35,105) {\footnotesize{$\Pra=10^{-3}$}}
\put (-15,70) {\footnotesize{$\displaystyle-\frac{3\pi}{8}$}}
\put (100,70)  {\footnotesize{$\displaystyle 0$}}
\put (-6,25) {\footnotesize{$\displaystyle\frac{\pi}{2}$}}
\put (98,25)  {\footnotesize{$\displaystyle\frac{5\pi}{8}$}}
\end{overpic} \\[5mm]
\begin{overpic}[height=0.24\columnwidth,tics=10]{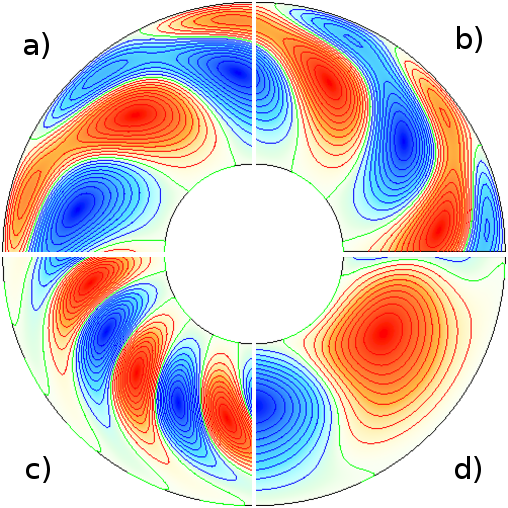}
\put (35,105) {\footnotesize{$\Pra=10^{-1}$}}
\put (-15,70) {\footnotesize{$\displaystyle-\frac{3\pi}{8}$}}
\put (100,70)  {\footnotesize{$\displaystyle 0$}}
\put (-6,25) {\footnotesize{$\displaystyle\frac{\pi}{2}$}}
\put (98,25)  {\footnotesize{$\displaystyle\frac{5\pi}{8}$}}
\end{overpic} &
\begin{overpic}[height=0.24\columnwidth,tics=10]{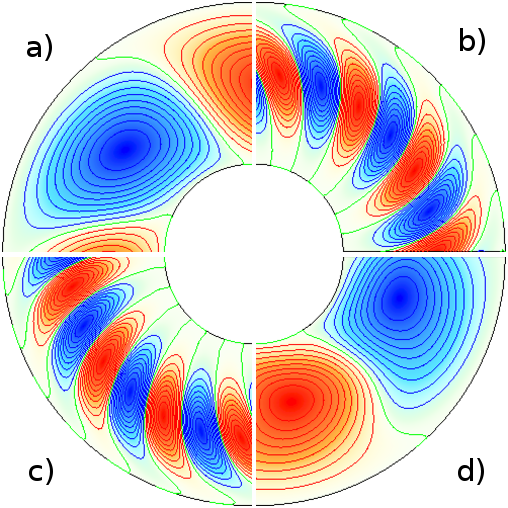}
\put (35,105) {\footnotesize{$\Pra=10^1$}}
\put (-15,70) {\footnotesize{$\displaystyle-\frac{3\pi}{8}$}}
\put (100,70)  {\footnotesize{$\displaystyle 0$}}
\put (-6,25) {\footnotesize{$\displaystyle\frac{\pi}{2}$}}
\put (98,25)  {\footnotesize{$\displaystyle\frac{5\pi}{8}$}}
\end{overpic} \\
\end{tabular}
\caption{Contour lines of the stream function $r^{-1} \upartial_\theta
\cal S$  of the flow in the equatorial plane at onset for four
representative values of the Prandtl number $\Pra$ and four
representative values of the mixing angle $\alpha$. Top left:
$\Pra=10^{-5}$; top right: $\Pra=10^{-3}$; bottom left: $\Pra=10^{-1}$;
bottom  right: $\Pra=10^1$. In each panel a) corresponds to
$\alpha=-3\pi/8$; b) to  $\alpha=0$; c) to $\alpha=\pi/2$; and d)
to $\alpha=5\pi/8$. In all cases $\Sc=25 \Pra$, $\tau = 10^4$ and
$\eta=0.35$. Clockwise and counter-clockwise eddies alternate. 
(Color online)
} 
\label{f:Pc.ggt.Pt.flowStruct}
\end{figure}
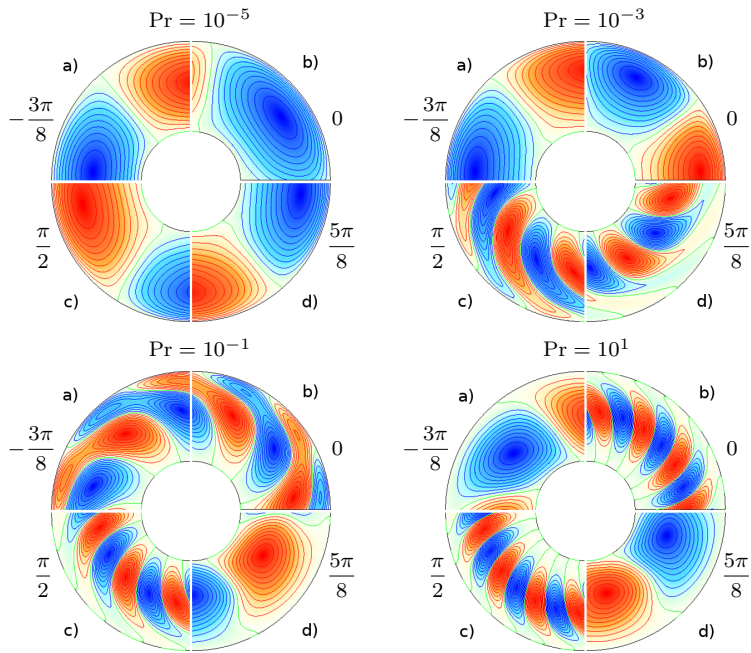

{Figure \ref{f:Pc.ggt.Pt.flowStruct} illustrates the corresponding
patterns of convection with the case $\Pra=10^{-3}$ plotted in the top
right panel and the case $\Pra=10^{-1}$ plotted in the lower left
panel.}

\subsubsection{Large Prandtl number case}
\label{largelargePr}
{The onset of convection for large values of the Prandtl
number, $\Pra>0.1$,} is illustrated in figure~\ref{f:Pc.ggt.Pt}(b) {where}
the drift rate, the most unstable wave number and  the critical
effective Rayleigh number {are plotted as a function of the
mixing angle $\alpha$.} As $\Pra$ is increased $\Ra$ and $m$ approach
limiting values and $\omega$ appears to tend to zero as expected from
the approximation~\eqref{highPasym}. Four different regimes
of convection can be identified. These are described in turn below
starting from negative values of $\alpha$. {We focus on the curve
$P=1$ in order to be able to quote specific parameter values.}

{For values of $\alpha$ smaller than approximately $-\pi/4$ motions are due to a diffusive
instability where the most unstable mode is an additional
double-diffusive mode as described in relation to
figures~\ref{f:IntroducingRcRtplane} and \ref{f:eigenModes}.} 
The gradient of the buoyancy profile $\Xi$ changes sign with respect
to {the gradients of} $T(r)$ and $C(r)$ and is shallower. 
{A similar situation} was analysed in section \ref{s:smallDepartures}
where an asymptote located at $\alpha=-\pi/4$ has now shifted to
$\alpha=-\pi/2$. The diffusion contribution to the buoyancy field
$\Psi$ is still dominated by the temperature but it is strongly
reduced since $\cos\alpha<1/2$. The buoyancy force, on the other hand, is
dominated by the concentration but acts in reversed direction pushing
regions of positive $\Psi$ down and regions of negative $\Psi$
up. This gives rise to  large scale ($m=3$), slow convection drifting
counter-clockwise as illustrated in figure~\ref{f:Pc.ggt.Pt} {(bottom
right plot, panel (a))}. The onset of convection occurs at lower values
of the effective Rayleigh number than for purely thermal convection.

{For values of $\alpha$ approximately between $-\pi/4$ and $\pi/4$ motions are due to
single-diffusive eigenmodes of purely thermal nature.} Indeed, when
values of $\alpha$ {are close to zero}
$\cos\alpha\approx 1-\alpha^2/2$ and $\sin\alpha\approx \alpha$ so
that the temperature remains the  only quantity being diffused in
equation~\eqref{eq:rhoWhenPt.ne.Pc} and it is also the main source of
buoyancy. The system behaves like a single-diffusive purely thermal
convection at $\Pra=10$ and exhibits values of the azimuthal wave
number, drift rate and effective Rayleigh number {as expected from}
equations \eqref{highPasym}, with the effective Rayleigh number
modified according to equation \eqref{eq:RaBreuer}. The spatial
pattern of the flow is illustrated in figure~\ref{f:Pc.ggt.Pt} {(bottom
right plot, panel (b))}. Abrupt transitions to neighbouring regimes occur at
both ends of this region, namely an abrupt transition to
diffusive-instability at $\alpha\approx -\pi/4$ and an abrupt
transition to chemical convection at $\alpha\approx\pi/12$.  

{For values of $\alpha$ approximately between $\pi/4$ and $\pi/2$ motions are due to
motions are due to
single-diffusive eigenmodes of purely chemical nature. These are
modes otherwise similar to the thermal convective modes of the regime 
to the left but characterised by a different value of the Prandtl
number -- the value of the Schmidt number is $\Sc=250$.}
This relatively large change in the values of the Prandtl
numbers results in a large abrupt jump in wave number to much smaller scales,
reduction of the drift rate by an order of magnitude and a discrete
change in slope of the $\Ra$ curve. Because these are
effectively single-diffusive modes the critical values are well
approximated by equations \eqref{highPasym} with $\Ra$ modified by equation
\eqref{eq:RaBreuer}.  The spatial pattern of the flow is illustrated
in figure~\ref{f:Pc.ggt.Pt} {(bottom right plot, panel (c))}. 

{For values of $\alpha$ greater than approximately $\pi/2$
 motions are due to a fingering
instability where the most unstable mode is an additional
double-diffusive mode as described in relation to figures~\ref{f:IntroducingRcRtplane} and \ref{f:eigenModes}.} 
For $\alpha>\pi/2$ the factor $\cos\alpha$ becomes 
negative and diffusion then acts to concentrate anomalies of $\Psi$
around the anomalies of $\Theta$ instead of dispersing them. However,
because $\Xi$ is now smaller than either of $T$ or $C$, there will be
less buoyancy produced by advection. 
This leads a relatively gradual transition between the purely chemical and the
fingering regime as illustrated in figure~\ref{f:Pc.ggt.Pt}(b) where a
continuous decrease in the value of $m$ towards a constant value of
$m=3$ and a continuous change in the slope of the $\Ra$ curve are
observed. Only the drift rate shows a small discrete jump to smaller
values. The spatial pattern of the flow is illustrated in
figure~\ref{f:Pc.ggt.Pt} {(bottom right plot, panel (d))}. 
The asymptote on {the far right side remains} at $3\pi/4$ as was the case in
discussed in relation to figure~\ref{f:Le.ap.1posd}.

\section{Dependence on the Coriolis number}
\label{s:tauDependence}
So far we have analysed cases at fixed {values of the} Coriolis
number, $\tau$. {In this section we investigate the dependence of
the onset of convection on this adimensional parameter.} In the case
of purely thermal convection the thermal Rayleigh number $\Rt$ {of 
columnar convection} is proportional to $\tau^{4/3}$ {as
approximated by} equation~\eqref{highPasym}. Here, we will show that   the
effective Rayleigh number for double-buoyant, double-diffusive convection obeys
the same scaling in regions where the lowest diffusion quantity dominates in
generating buoyancy including in the regions where both Rayleigh
numbers are positive. {We will also establish numerically an
approximate scaling valid in the regions of diffusive and fingering regimes.}

\begin{figure}
{\footnotesize (a) \hspace{0.5\textwidth} (b)} \\
\subfigure{\resizebox*{0.498\textwidth}{!}%
{\includegraphics{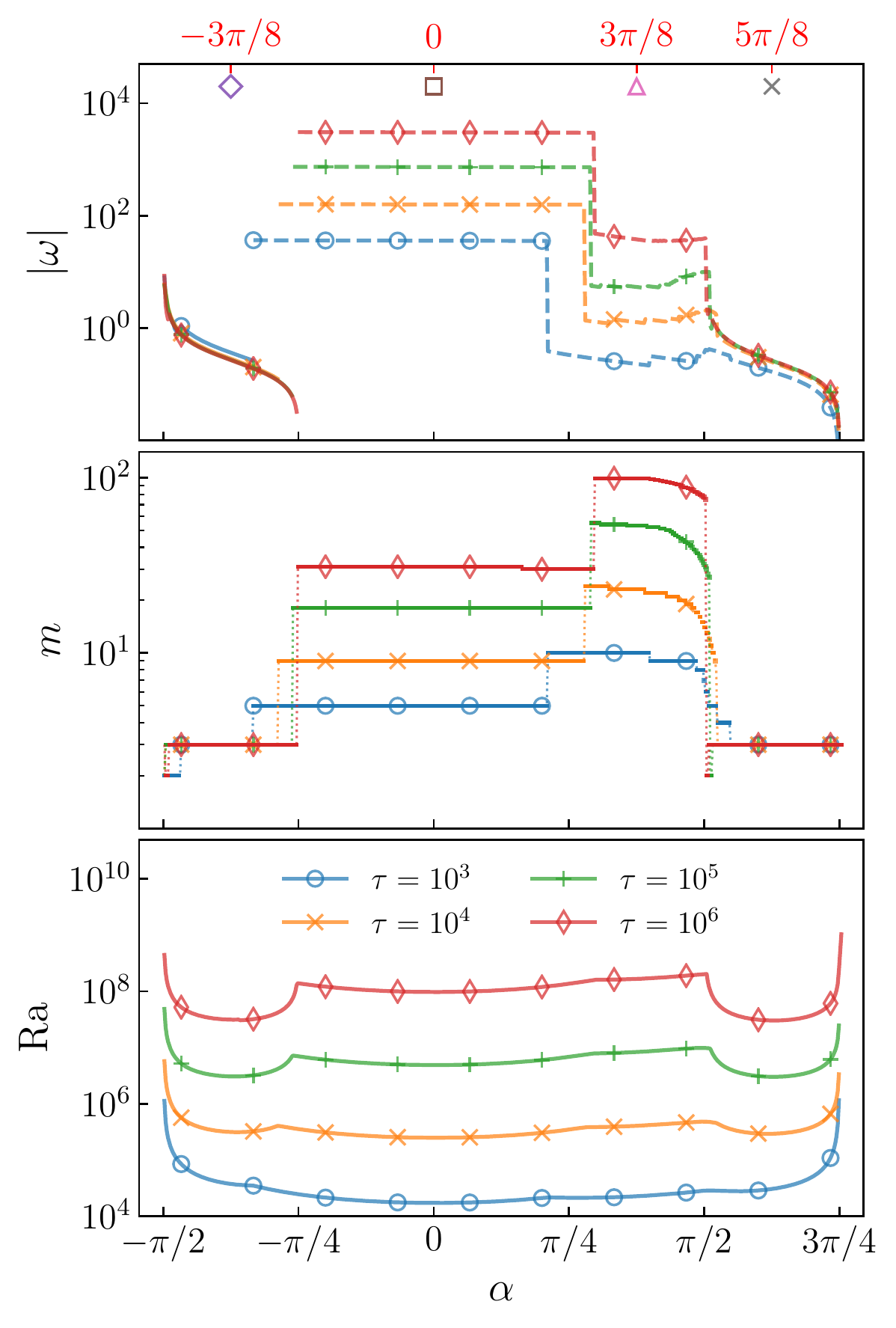}}}
\subfigure{\resizebox*{0.50\textwidth}{!}%
	{\includegraphics{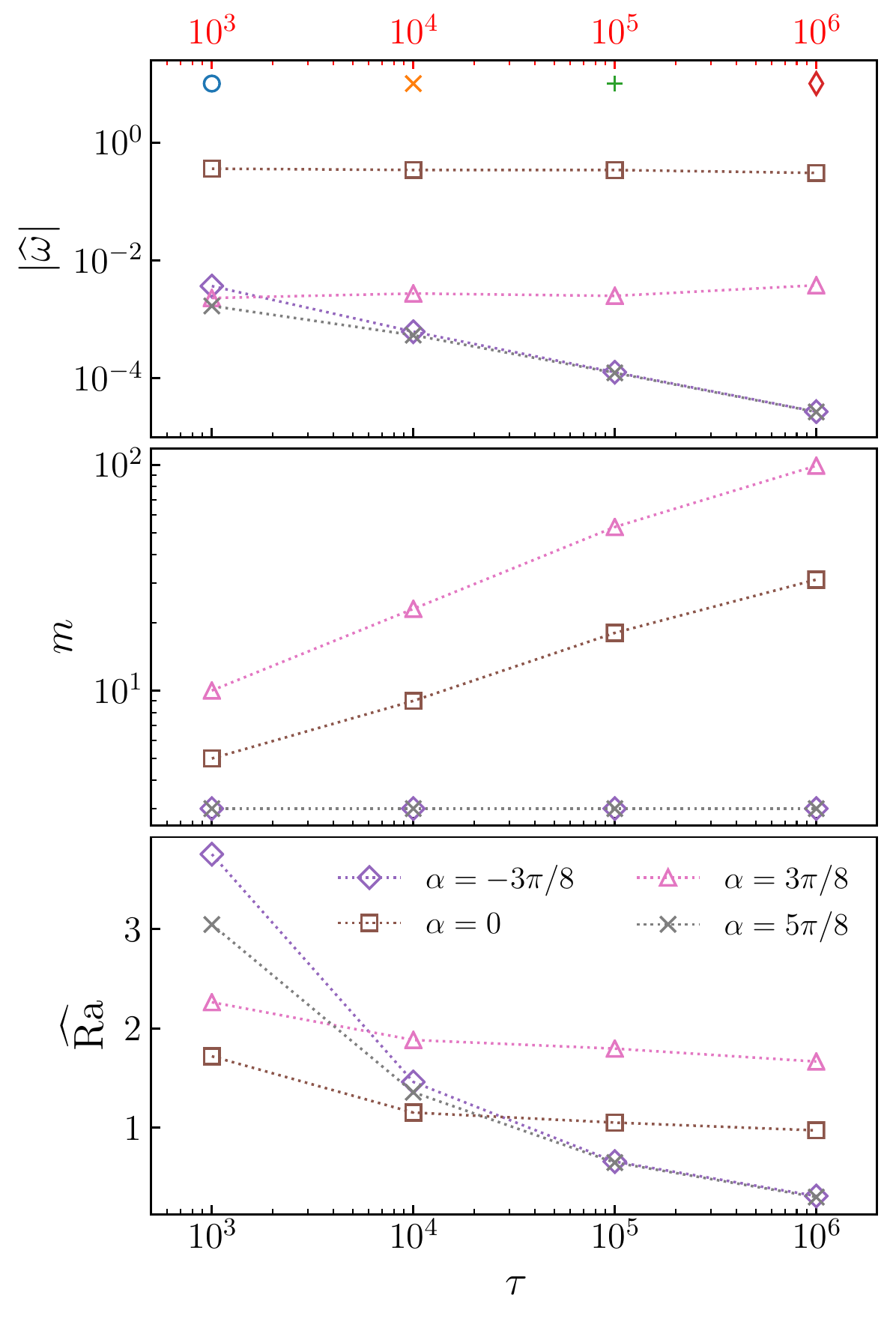}}}
\caption{{
Critical parameter values for the onset of convection at $\Pra=1$,
$\Sc=100$, $\eta=0.35$ as a function of the Coriolis number $\tau$ and
the mixing angle $\alpha$. (a) Critical Rayleigh number $\Ra$ (bottom
panel), most unstable wave-number $m$ (middle panel), and drift-rate
amplitude $|\omega|$ (top panel) as a function of $\alpha$ for values
of $\tau$ as specified in the legend. Negative values of $\omega$ are
indicated by dashed lines and positive values are indicated by solid
lines.  Note, red ticks with symbols on the uppermost x-axis denote
selected values of $\alpha$ at which the curves in (b) are sampled.
(b) Scaled critical Rayleigh number $\widehat{\Ra}=\Ra\tau^{-4/3}$
(bottom
panel), most unstable wave-number $m$ (middle panel), and scaled drift-rate
amplitude $\widehat{\omega}=|\omega|\tau^{-2/3}$ (top panel) as a
function of $\tau$ for values of $\alpha$ as specified in the legend.
Note, red ticks with symbols on the uppermost x-axis denote
selected values of $\tau$ at which the curves in (a) are sampled.
}
(Color online)
}
 \label{f:dependenceTau}
\end{figure}

Figure~\ref{f:dependenceTau}(a) {shows critical parameter values
for the onset of convection as a function of $\alpha$ for selected
values of $\tau$ and}
for fixed valued of the Prandtl numbers $\Pra=1$ and $\Sc=100$ and
{shell aspect ratio $\eta=0.35$}. Since both Prandtl numbers are
larger than unity {the discussion presented in Section
  \ref{largelargePr} is relevant}. {As the mixing angle $\alpha$
is varied for a fixed value of $\tau$ the four convective regimes identified in
the later section are evident from} the `boat-shaped' plots which
appear similar to those in {figure~\ref{f:Pc.ggt.Pt}(b).} 
We observe a region symmetric around $\alpha=0$ where instability is
due to purely thermal eigenmodes since buoyancy has a predominantly
thermal component. 
The region where buoyancy has a predominantly chemical component
occupies {a range of mixing angle values} approximately between
$\alpha=\pi/4$ and $\alpha=\pi/2$.  
Both purely thermal and purely compositional modes {take the form of
columnar   convection} featuring a cartridge belt of $z$-aligned columns
centred at mid shell as expected at these relatively large values of
the Prandtl numbers. The transition between the two regimes is
characterised by large abrupt jumps in the values of the drift rate
$omega$ and the wave number $m$ due to the relative difference in the
values of $\Pra$ and $Sc$.
{The single-diffusive regimes are flanked} by regions of very
large scale convection appear approximately in the octants $\alpha\in[-\pi/2, -\pi/4]$ and
$\alpha\in[\pi/2, 3\pi/4]$ {corresponding to diffusive and fingering
instabilities, respectively}. These regions are characterised by 
Rayleigh numbers of opposite signs. This shape and convective structure 
were discussed in Section~\ref{s:largePrDifferences}.

Figure~\ref{f:dependenceTau}(b) shows {explicitly the dependence of the
critical parameter values for the onset of convection on the Coriolis
number $\tau$ for several selected values of $\alpha$ sampling the
four distinct convective regimes.} {In order to compare trends with the
approximations \eqref{highPasym} of \citet{Yano1992}}
the scaled quantities $\widehat \Ra=\Ra\tau^{-4/3}$ and $\widehat
\omega=\omega\tau^{-2/3}$  have been actually plotted.
The curves for all values of $\alpha$ sampled in
figure~\ref{f:dependenceTau}(b) which belong to the regions of
purely-thermal or purely-chemical instability follow the 
scaling of approximations \eqref{highPasym} namely, 
$\Ra \propto \tau^{4/3}$ and $\omega\propto \tau^{2/3}$ as
demonstrated by the curves becoming horizontal for the larger values
of $\tau$. Although the wave number $m$ is not scaled in this figure
it also also follows a thermal-type scaling $m\propto \tau^{1/3}$
that we expect from equations \eqref{highPasym}. The purely
For instance, a purely thermal case occurs when $\alpha=0$ (filled
triangles) so that $\Ra=\Rt$ and in this case the particular curves in
figure~\ref{f:dependenceTau}(b) correspond to the curves for $\Pr=1$
in figure~3 of \citep{ArdesEtAl97}, although there is a slight
discrepancy as $\eta=0.4$ in the latter work. 
Similarly, when $\alpha=\pi/2$ (filled circles) convection is purely
chemical and if the values of  $\Rc$, $\Sc$ are used the asymptotic
trends for columnar convection cn be reproduced.
A different scaling emerges in the two essentially double-diffusive
regions sampled by points $\alpha=-3\pi/8$ and $\alpha=5\pi/8$. These
are the regions previously described as arising from large
discrepancies in Prandtl numbers. The most unstable azimuthal wave
number in these regions seems to be independent of the Coriolis
number, $\tau$, and the critical effective Rayleigh number is now $\Ra\propto\tau$.
Note that, because these new regions obey a shallower scaling of $\Ra$ with 
$\tau$, as this parameter grows, those large scale regions will have a
lower critical value of  $\Ra$ when compared to purely thermal or
chemical convection. 

Significant deviations from the scaling discussed {of
equation~\eqref{highPasym}} seem to occur ony for small values of the
Coriolis number. {This is expected for two reasons. Firstly, 
at small $\tau$ the end of the range of validity of the
approximation~\eqref{highPasym} is approached, and secondly, inertial 
effects start to play a role when $\tau\Pra = O(10^3)$,
\citep{Simitev2003,BusseSimitev2004}.}  

\section{Dependence on the shell thickness}
\label{s:etaDependence}
{In this last section the dependence} on the shell thickness is
explored. The Earth's inner core grows due to secular cooling and
different convective regimes may occur at various moments in geological time. 

\begin{figure}
{\footnotesize (a) \hspace{0.5\textwidth} (b)} \\
\subfigure{\resizebox*{0.499\textwidth}{!}%
	{\includegraphics{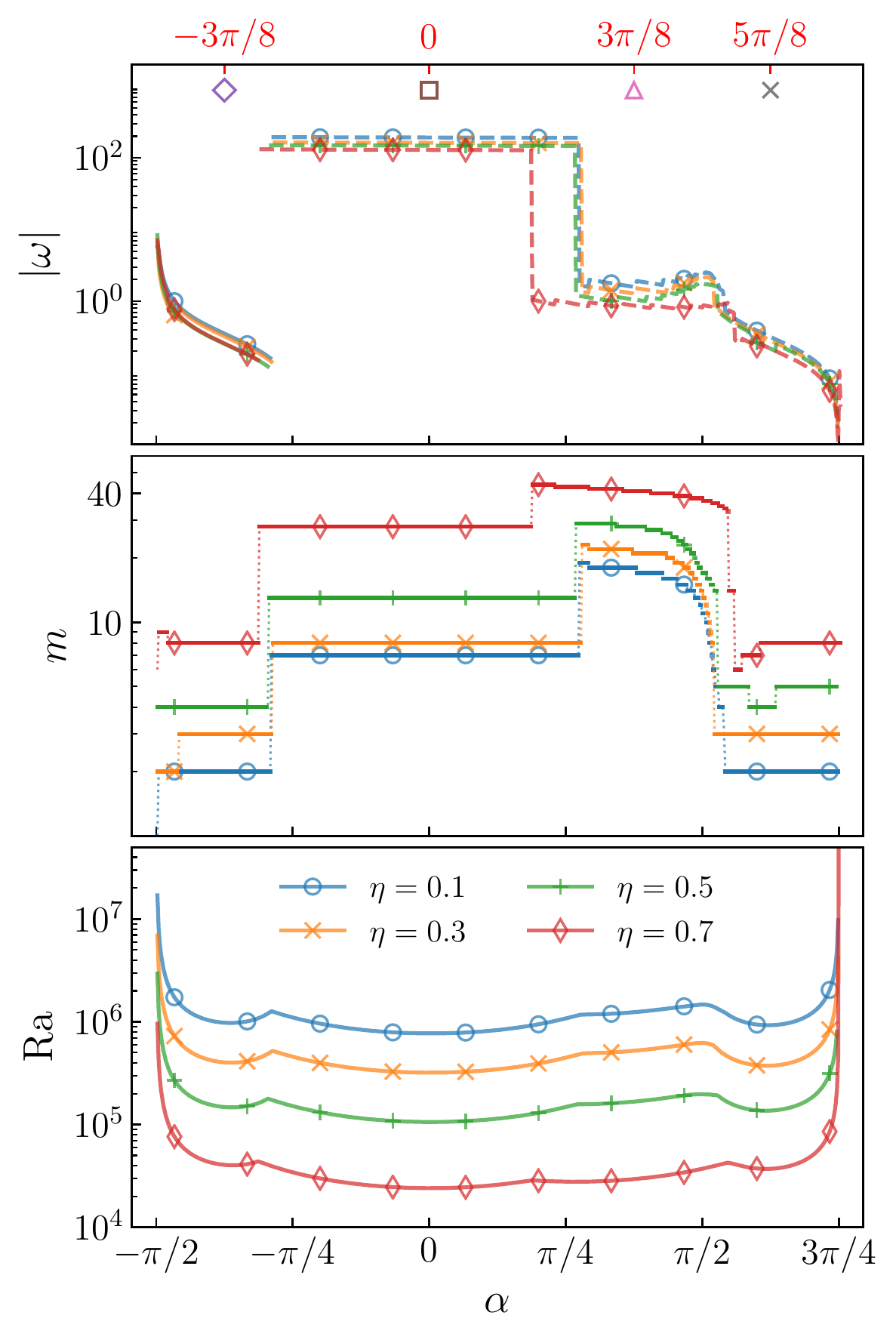}}}
\subfigure{\resizebox*{0.50\textwidth}{!}%
	{\includegraphics{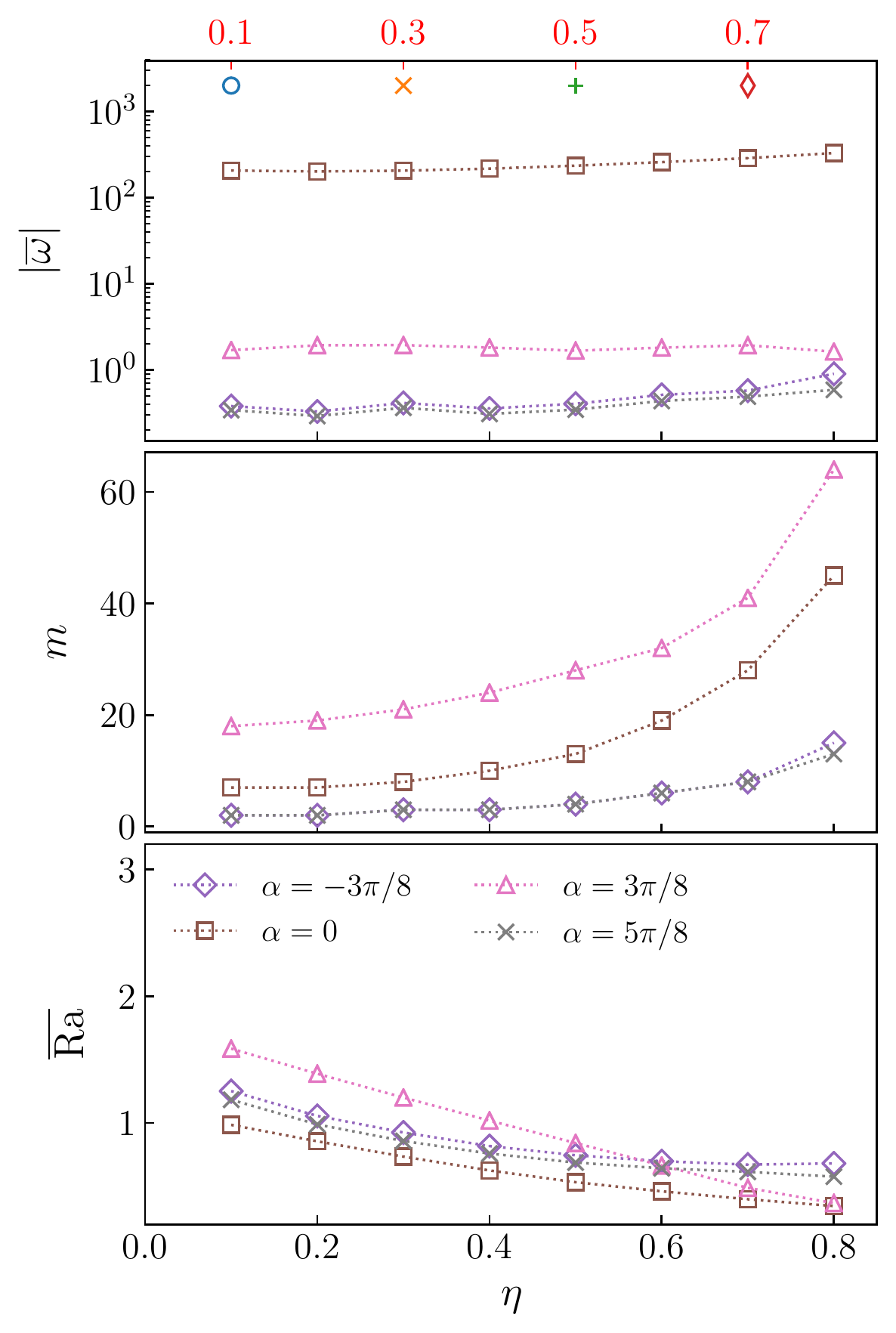}}}
 \caption{
{
Critical parameter values for the onset of convection at $\Pra=1$,
$\Sc=100$, $\tau=10^4$ as a function of the shell aspect ratio $\eta$ and
the mixing angle $\alpha$. (a) Critical Rayleigh number $\Ra$ (bottom
panel), most unstable wave-number $m$ (middle panel), and drift-rate
amplitude $|\omega|$ (top panel) as a function of $\alpha$ for values
of $\eta$ as specified in the legend. Negative values of $\omega$ are
indicated by dashed lines and positive values are indicated by solid
lines.  Note, red ticks with symbols on the uppermost x-axis denote
selected values of $\alpha$ at which the curves in (b) are sampled.
(b) Scaled critical Rayleigh number $\overline{\mathrm{Ra}}=\mathrm{Ra}(1-\eta)^{-7/3}10^{-6}$
(bottom
panel), most unstable wave-number $m$ (middle panel), and scaled drift-rate
amplitude $\overline{\omega}=\omega(1-\eta)^{-2/3}$ (top panel) as a
function of $\eta$ for values of $\alpha$ as specified in the legend.
Note, red ticks with symbols on the uppermost x-axis denote
selected values of $\eta$ at which the curves in (a) are sampled.
}
(Color online)
}   
 \label{f:dependenceEta}
\end{figure}

{Similarly to past sections, we keep all but two} of the parameters at fixed
values, in particular $\Pra=1$, $\Sc=100$ and $\tau=10^4$, while 
varying the mixing angle $\alpha$ and the parameter of interest, in this
case the aspect ratio $\eta$.
Te critical values of the Rayleigh number, the most unstable wave
number and the drift rate are plotted as functions of the two variable
parameters in figure~\ref{f:dependenceEta}. 
As the value of the mixing angle $\alpha$ is varied in
figure~\ref{f:dependenceEta}(a) the four convective regimes familiar
from the past two sections are observed at all values of $\eta$.
The boundaries between these regimes vary little with $\eta$ -- the
boundary between the diffusive and the thermal columnar regime is
located near $-\pi/4$, the boundary between the  thermal columnar
regime and the chemical columnar regime is located near $\pi/4$, and
the boundary between the chemical columnar regime and the fingering
regime is located near $\pi/2$, all of them sifting to smaller values
with increasing $\eta$. In order to investigate the dependence on $\eta$
directly and to compare trends with the approximation
\eqref{highPasym}, we plot in figure~\ref{f:dependenceEta}(b) the
scaled quantities $\overline\omega=\omega(1-\eta)^{-2/3}$, $m$ and
$\overline\Ra=\Ra(1-\eta)^{-7/3}$ as function of $\eta$. The
approximation appears to hold well for thick spherical
shells $\eta \in[0.1,0.5]$ but fails for thin shells where the effects of
the curved boundaries start to dominate and columnar structures cannot
easily form. The approximation for $\Ra$, naturally, also fails for
values of $\alpha$ that are not in the columnar regimes. 
In these regimes when the system is stabilised by one quantity but
destabilised by the other the most unstable wave number seems to be
proportional to the mean radius and the effective Rayleigh number, to
the square of the radius at 44\% of the shell.

\section{Discussion and conclusions}
\label{s:conclusions}
In this paper we identify deviations of the onset of doubly buoyant,
doubly diffusive spherical rotating convection from the purely thermal case. We 
introduce a new representation for the thermal and compositional Rayleigh 
numbers that makes it possible to define a strictly positive pre-factor for the 
buoyancy. This pre-factor is the effective Rayleigh number $\Ra$
that, in this representation, has a unique critical value {at
fixed values} of all other parameters. 
{In contrast, when either of $\Rt$ or $\Rc$ serves as a critical parameter it may
take negative values and the curve of marginal stability may be
single, double- or triple-valued. In our new representation,} the
trade-off between thermal and compositional buoyancy is determined by
a mixing angle $\alpha$ which takes positive values in the right half
of the $\Rt-\Rc$ plane. 

{Four different regimes of convection are identified depending on the
values of the mixing angle $\alpha$ and the Prandtl and Schmidt numbers.}
In the case when the values of the Prandtl number and the Schmidt
number are equal {convection} behaves like {in} a
single-diffusive, one-buoyancy system for all values of the mixing angle. In the $\Rt-\Rc$ plane the
critical curve is a straight line with slope $-1$. In the $\Ra-\alpha$
plane the marginal curve behaves like $1/(\cos\alpha + \sin\alpha)$
and approaches vertical asymptotes at $-\pi/4$ and $3\pi/4$.
When the values of the  Prandtl number and the Schmidt number are
slightly different from each other a fingering and a diffusive
convection regimes emerge due to the  different rates of diffusion of
the two components of the fluid mixture. A small-deviations
approximation allows us to gain some insight into the nature of these
regimes. When the values of the Prandtl and the Schmidt numbers are
significantly different from each other an additional transition
between a thermal columnar and a thermal chemical regime occurs if the
values of $\Pra$ are large. For small values of $\Pra$ both the thermal
and the chemical modes are in an inertial regime without a clearly
identifiable transition. {In more detail,} when both Prandtl
numbers are below 0.1 very large scale inertial convection dominates
independently of $\alpha$. When both Prandtl numbers are above 0.1,
rotating convection takes place with a variety of wave numbers
depending on $\alpha$.  
{In regions of parameter space where} $\Rt$ or $\Rc$ are negative, only very
large scale convection seems to be possible. In terms of mixing angle
these regions correspond to $\alpha<0$ or $\alpha>\pi/2$ and the modes
seem related to the diffusive and fingering regimes of
in thermohaline systems. Convection
is not possible when the quantity with the {smaller} Prandtl
number and a negative Rayleigh number dominates buoyancy. When the
quantity with the {smaller} Prandtl number and a negative Rayleigh
number does not dominate the buoyancy budget, convection is very large
scale and has a negative drift rate {(prograde)}. This structure of
convection appears to be qualitatively preserved for all values of the
Coriolis number $\tau$ and the shell thickness $\eta$ investigated.

Most of this study was carried out for and relatively large values of
Prandtl and Schmidt numbers and {for moderate values
of the Coriolis number $\tau$.  This choice offers a reasonable
trade-off between computational expense and the possibility to
compare the results to known trends for columnar convection}
\citep{Busse1970,Yano1992}.
{Both thermal and  chemical columnar modes which are observed for $\tau\Pra > O(10^3)$}
appear to satisfy the trends of the approximation given
by equation~\eqref{highPasym}, as long as the regime of large scale
convection with either $\Rt<0$, $\Rc>0$ or $\Rt>0$, $\Rc<0$ is not
considered.
In those regions, $\Ra\propto\tau$, the preferred
azimuthal wave number $m$ shows little variation with respect to $\tau$
and the drift rates scales as $|\omega|\propto \tau^{2/3}$. The
columnar scaling is also {observed} when varying $\eta$ but only
for relatively thick shells $\eta\in[0.1,0.5]$ beyond which the
approximation given by equation~\eqref{highPasym} starts to break
down. Again this thermal scaling is only correct as long as the regime
of large scale convection with either $\Rt<0$, $\Rc>0$ or $\Rt>0$,
$\Rc<0$ is not considered.  
 
\section*{Acknowledgements} 
This work was supported by the Leverhulme Trust [grant number RPG-2012-600].


\end{document}